\documentclass[a4paper,11pt]{article}
\usepackage{jcappub}
\usepackage{aas_macros}
\usepackage{lineno}
\usepackage{orcidlink}
\usepackage{bm}

\newcommand{\numMocks}{20}

\newcounter{enumlast}
\newcommand{\lya}{Ly\ensuremath{\alpha}}
\newcommand{\poned}{$P_{\mathrm{1D}}$}
\newcommand{\skm}{s~km$^{-1}$}
\newcommand{\kms}{km~s$^{-1}$}

\newcommand{\pk}{$P_{\mathrm{1D}}$}

\newcommand{\qq}{\texttt{quickquasars}}

\title{\boldmath DESI DR1 Ly$\alpha$ 1D power spectrum: Validation of estimators}

\author[1,2,3]{{Naim~G\" oksel Kara\c{c}ayl{\i}}\orcidlink{0000-0001-7336-8912},}
\author[4]{{Corentin Ravoux}\orcidlink{0000-0002-3500-6635},}
\author[1,2]{{Paul Martini}\orcidlink{0000-0002-4279-4182},}
\author[5]{{Jean-Marc Le~Goff},}
\author[5]{{Eric Armengaud}\orcidlink{0000-0001-7600-5148},}
\author[5]{{M.~Abdul-Karim}\orcidlink{0009-0000-7133-142X},}
\author[6]{{J.~Aguilar},}
\author[7]{{S.~Ahlen}\orcidlink{0000-0001-6098-7247},}
\author[6]{{A.~Anand}\orcidlink{0000-0003-2923-1585},}
\author[8]{{S.~BenZvi}\orcidlink{0000-0001-5537-4710},}
\author[9,10]{{D.~Bianchi}\orcidlink{0000-0001-9712-0006},}
\author[11]{{D.~Brooks},}
\author[6]{{T.~Claybaugh},}
\author[6,12]{{A.~Cuceu}\orcidlink{0000-0002-2169-0595},}
\author[13]{{A.~de la Macorra}\orcidlink{0000-0002-1769-1640},}
\author[14,15]{{Biprateep~Dey}\orcidlink{0000-0002-5665-7912},}
\author[11]{{P.~Doel},}
\author[6,16]{{S.~Ferraro}\orcidlink{0000-0003-4992-7854},}
\author[17]{{A.~Font-Ribera}\orcidlink{0000-0002-3033-7312},}
\author[18,19]{{J.~E.~Forero-Romero}\orcidlink{0000-0002-2890-3725},}
\author[20,21,22]{{E.~Gaztañaga}\orcidlink{0000-0001-9632-0815},}
\author[6,23]{{S.~Gontcho A Gontcho}\orcidlink{0000-0003-3142-233X},}
\author[24]{{G.~Gutierrez},}
\author[25,5]{{H.~K.~Herrera-Alcantar}\orcidlink{0000-0002-9136-9609},}
\author[1,3]{{K.~Honscheid}\orcidlink{0000-0002-6550-2023},}
\author[26]{{M.~Ishak}\orcidlink{0000-0002-6024-466X},}
\author[17]{{J.~Jimenez}\orcidlink{0000-0001-8528-3473},}
\author[27]{{R.~Joyce}\orcidlink{0000-0003-0201-5241},}
\author[28]{{D.~Kirkby}\orcidlink{0000-0002-8828-5463},}
\author[6]{{T.~Kisner}\orcidlink{0000-0003-3510-7134},}
\author[6]{{A.~Kremin}\orcidlink{0000-0001-6356-7424},}
\author[11]{{O.~Lahav},}
\author[6]{{M.~Landriau}\orcidlink{0000-0003-1838-8528},}
\author[29]{{L.~Le~Guillou}\orcidlink{0000-0001-7178-8868},}
\author[30,17]{{M.~Manera}\orcidlink{0000-0003-4962-8934},}
\author[27]{{A.~Meisner}\orcidlink{0000-0002-1125-7384},}
\author[31,17]{{R.~Miquel},}
\author[21]{{S.~Nadathur}\orcidlink{0000-0001-9070-3102},}
\author[32,33]{{G.~Niz}\orcidlink{0000-0002-1544-8946},}
\author[5,6]{{N.~Palanque-Delabrouille}\orcidlink{0000-0003-3188-784X},}
\author[34,35,36]{{W.~J.~Percival}\orcidlink{0000-0002-0644-5727},}
\author[6,37,16]{{C.~Poppett},}
\author[38]{{F.~Prada}\orcidlink{0000-0001-7145-8674},}
\author[39]{{I.~P\'erez-R\`afols}\orcidlink{0000-0001-6979-0125},}
\author[40]{{G.~Rossi},}
\author[41]{{E.~Sanchez}\orcidlink{0000-0002-9646-8198},}
\author[6]{{D.~Schlegel},}
\author[42,43]{{M.~Schubnell},}
\author[44]{{H.~Seo}\orcidlink{0000-0002-6588-3508},}
\author[6]{{J.~Silber}\orcidlink{0000-0002-3461-0320},}
\author[27]{{D.~Sprayberry},}
\author[5]{{T.~Tan}\orcidlink{0000-0001-8289-1481},}
\author[43]{{G.~Tarl\'{e}}\orcidlink{0000-0003-1704-0781},}
\author[45,46]{{M.~Walther}\orcidlink{0000-0002-1748-3745},}
\author[27]{{B.~A.~Weaver},}
\author[47]{{H.~Zou}\orcidlink{0000-0002-6684-3997},}

\affiliation[1]{Center for Cosmology and AstroParticle Physics, The Ohio State University, 191 West Woodruff Avenue, Columbus, OH 43210, USA}
\affiliation[2]{Department of Astronomy, The Ohio State University, 4055 McPherson Laboratory, 140 W 18th Avenue, Columbus, OH 43210, USA}
\affiliation[3]{Department of Physics, The Ohio State University, 191 West Woodruff Avenue, Columbus, OH 43210, USA}
\affiliation[4]{Universit\'{e} Clermont-Auvergne, CNRS, LPCA, 63000 Clermont-Ferrand, France}
\affiliation[5]{IRFU, CEA, Universit\'{e} Paris-Saclay, F-91191 Gif-sur-Yvette, France}
\affiliation{Remaining affiliations are in Appendix \ref{sec:affiliations}}
\emailAdd{karacayli.1@osu.edu}
\emailAdd{corentin.ravoux@clermont.in2p3.fr}

\abstract{
The Data Release 1 (DR1) of the Dark Energy Spectroscopic Instrument (DESI) is the largest sample to date for small-scale Ly$\alpha$ forest cosmology, accessed through its one-dimensional power spectrum ($P_{\mathrm{1D}}$). The Ly$\alpha$ forest $P_{\mathrm{1D}}$ is extracted from quasar spectra that are highly inhomogeneous (both in wavelength and between quasars) in noise properties due to intrinsic properties of the quasar, atmospheric and astrophysical contamination, and also sensitive to low-level details of the spectral extraction pipeline. We employ two estimators in DR1 analysis to measure $P_{\mathrm{1D}}$: the optimal estimator and the fast Fourier transform (FFT) estimator. To ensure robustness of our DR1 measurements, we validate these two power spectrum and covariance matrix estimation methodologies against the challenging aspects of the data. First, using a set of 20 synthetic 1D realizations of DR1, we derive the masking bias corrections needed for the FFT estimator and the continuum fitting bias needed for both estimators. We demonstrate that both estimators, including their covariances, are unbiased with these corrections using the Kolmogorov–Smirnov test. Second, we substantially extend our previous suite of CCD image simulations to include 675,000 quasars, allowing us to accurately quantify the pipeline's performance. This set of simulations reveals biases at the highest $k$ values, corresponding to a resolution error of a few percent. We base the resolution systematics error budget of DR1 $P_{\mathrm{1D}}$ on these values, but do not derive corrections from them since the simulation fidelity is insufficient for precise corrections.
}

\begin{document}
\maketitle
\flushbottom


\section{Introduction\label{sec:intro}}
The 1D flux power spectrum of the \lya\ forest (\poned) is rich in information on small-scale physics (below one Mpc). It has been used to constrain the thermal state of the intergalactic medium (IGM) \cite{boeraRevealingThermal2019, waltherNewConstraintsIGM2019, villasenorThermalHistory2022}, the amplitude and the slope of the linear matter power spectrum at $k\approx1~$Mpc$^{-1}$ \cite{croftRecoveryPowerSpectrum1998, mcdonald_observed_2000, pedersenEmulator2021}, the primordial power spectrum \cite{vielPrimordialPowerSpectrumLya2004}, the sum of neutrino masses \cite{croftNeutrinoMassLyaForest1999, palanqueDelabrouilleNeutrinoMass2015, yecheNeutrinoMassesXQ2017}, and the nature of dark matter \cite{narayananWDMLyaForest2000, seljakSterileNeutrinosDM2006, wangLyaDecayingDM2013, irsicFuzzyDMfromLya2017, boyarskyLyaWDM2009, vielWarmDarkMatter2013, baurLyaCoolWDM2016, irsicConstraintsWDM2017, villasenorWarmDarkMatter2023}. The Dark Energy Spectroscopic Instrument's (DESI) \cite{leviDESIExperimentWhitepaper2013, desicollaborationDESIExperimentPart2016} now-public Data Release 1 (DR1) \cite{desiKp2DataRelease12024} has over 300,000 \lya\ quasars, which is the largest sample to date for \poned\ measurements by a factor of 1.7 compared to the Extended Baryon Oscillation Spectroscopic Survey (eBOSS) \cite{dawsonSDSSeBOSSEarlyData2016}. This substantial increase in data size necessitates a thorough investigation of \poned\ and covariance matrix estimation techniques and spectroscopic processing to ensure robust cosmological inference.

This work consolidates the validation tests performed for the two estimators used in DESI DR1 \poned\ measurement --- the Fast Fourier Transform (FFT) estimator presented in ref.~\cite{ravouxFFTP1dDesiDr12024} and the quadratic maximum likelihood estimator (QMLE), also known as the optimal estimator, presented in ref.~\cite{karacayliQmleP1dDesiDr12024}, which differ at the level of their treatment of the data. The optimal estimator weights pixels using a covariance matrix constructed from the pipeline noise estimates and \lya\ forest correlations, which enables this estimator to be robust against gaps in the spectra that arise from masking contamination from unrelated physical and instrumental effects. The Fourier transform is significantly faster than the inverse covariance weighting of the optimal estimator, but requires simplifying assumptions in the treatment of noise, resolution, and biases due to masking, especially of atmospheric lines.

It is further critical to ensure that the estimated covariance matrix of \poned\ accurately and conservatively reflects the measurement errors. The FFT technique has a newly developed covariance matrix estimator presented in ref.~\cite{ravouxFFTP1dDesiDr12024} that we summarize and validate in this paper.
The optimal estimator produces a Gaussian covariance matrix that underestimates the errors and needs to be complemented with non-parametric bootstrap techniques \cite{mcdonaldLyUpalphaForest2006, karacayliOptimal1dDesiEdr2023}. The bootstrap covariance matrix is noisy and numerically unstable for redshift bins with low statistics. We develop and validate a new method for combining Gaussian and noisy bootstrap covariance estimates to ensure smoothness and prevent underestimation of errors.

We produce synthetic 1D \lya\ forest data using a log-normal transformation to validate the statistical robustness and identify any needed bias corrections \cite{karacayliOptimal1DLy2020}. These \lya\ transmission fields do not have any instrumental and astrophysical effects, which are added via DESI's simulation package. We analyze \numMocks\ realizations of DR1 and estimate a \poned\ and a covariance matrix for each realization. Since these mocks are inherently random, the conclusions are interpreted probabilistically using the Kolmogorov–Smirnov test. The corrections derived from these mocks are applied to the measurements in refs.~\cite{karacayliQmleP1dDesiDr12024, ravouxFFTP1dDesiDr12024}.

Since \lya\ forest measurements, and \poned\ specifically, are highly sensitive to the details of spectroscopic data processing, we also investigate the spectral extraction pipeline. The primary concern with spectral extraction is the accuracy of the resolution matrix, as errors increase quadratically with $k$ values, biasing the small-scale information. We test the accuracy of the resolution matrix by simulating 2D CCD images, which are extracted using DESI's spectroscopic data processing pipeline \cite{boltonSpectroPerfectionismAlgorithmicFramework2010, guySpectroscopicDataProcessingPipeline2022}. This procedure has been computationally expensive; however, after significant improvements to the pipeline, the images can now be efficiently processed using GPUs at a fraction of the cost, achieving a speedup of over a factor of 20. This enables us to match the sample size of DR1 by expanding our DESI early data release (EDR) simulated CCD images by a factor of 15 \cite{earlyDataRelease2023, karacayliOptimal1dDesiEdr2023}. These CCD image simulations help us quantify the systematic error budget in resolution, as we have done in refs.~\cite{karacayliQmleP1dDesiDr12024, ravouxFFTP1dDesiDr12024}. The simulations can also enable us to quantify any biases introduced by the extraction pipeline and serve as a means to test improvements in pipeline performance.
However, simulation fidelity needs to be improved before these results are used as bias corrections.

The outline of this paper is as follows. We describe the two types of simulated data in section~\ref{sec:synth_data}. Our methodology is presented in section~\ref{sec:methods}, which provides details on the \poned\ and covariance matrix estimators, as well as the quasar continuum fitting algorithm and the Kolmogorov–Smirnov test. The optimal estimator is validated in section~\ref{sec:qmle_val}, and the FFT estimator is validated in section~\ref{sec:fft_val}. We estimate the accuracy of the spectroscopic data processing and resolution matrix in section~\ref{sec:ccd_resolution_val}. The limitations of our analysis and directions for future work are discussed in section~\ref{sec:discuss}.

\section{Synthetic data\label{sec:synth_data}}
We follow the same procedure presented in the DESI early data release measurement to generate synthetic data~\cite{karacayliOptimal1dDesiEdr2023}. This procedure consists of two main components: generating 1D transmission fields with forest fluctuations and simulating the DESI instrument. 

The 1D transmission fields are created using the lognormal model in ref.~\cite{karacayliOptimal1DLy2020}, which we summarize briefly here. We first generate Gaussian random fields that are correlated along the line of sight, with an input power spectrum tuned to approximately match the desired flux power spectrum (these fields are uncorrelated between different lines of sight). Then, they are multiplied by the appropriate redshift evolution functions and are applied a lognormal transformation to obtain the column density, $n(z)$, and optical depth, $\tau(z)$, fields. The lognormal transform introduces nonlinearity and ensures that $n(z)$ and $\tau(z)$ are non-negative \cite{biAlternativeModelLyalpha1992, mcdonaldLyUpalphaForest2006}. Finally, the transmission field is defined as $F(z) = \mathrm{e}^{-\tau(z)}$. We generate these transmission fields on a linear wavelength grid with a 0.2\,\AA\ spacing, without any resolution or noise effects.

Simulating the DESI instrument also encapsulates creating a diverse set of quasar spectra, whose continua are generated from a broken power law with emission lines using the \texttt{simqso} package \cite{simqso} as a part of the \texttt{desisim} package.\footnote{\url{https://github.com/desihub/desisim}} We have two methods to simulate the DESI instrument.

The first method uses the \qq\ program in the same package. DESI observations are simulated using the \texttt{specsim}\footnote{\url{https://github.com/desihub/specsim}} package \cite{kirkbySpecsim2021} for quick simulations of the fiber spectrograph response (see ref.~\cite{herrera-alcantarDESIQuickquasars2023} for a detailed description of \qq\ mocks).
This program (after generating random quasar spectra) simulates sky and instrumental noise, and incorporates wavelength-dependent camera resolution, although it does not validate the computationally expensive spectral extraction process.

The second method uses CCD image simulations to validate the spectro-perfectionism resolution matrix \cite{boltonSpectroPerfectionismAlgorithmicFramework2010} and validate the complete spectral extraction process. This method projects mock quasar spectra (generated using \texttt{simqso}) onto two-dimensional images that simulate DESI raw data at the CCD pixel level with the \texttt{desisim} package. We then process these simulated images with the same spectroscopic reduction pipeline employed for actual DESI observations \cite{guySpectroscopicDataProcessingPipeline2022}. This approach is more computationally expensive than the mocks described above.

\subsection{Quickquasars mocks}
For these mocks, the quasar diversity, DESI instrument, and the sky are simulated through a program called \qq\ in the \texttt{desisim} package.
This program randomly generates quasar spectra from a broken power law with emission lines, convolves the spectra with the wavelength-dependent camera resolution for each spectrograph, adds noise appropriate to a given flux and exposure time, and resamples the resulting simulated spectra onto the output DESI wavelength grid of $\Delta\lambda_\mathrm{DESI}=0.8$\,\AA\ per pixel. The source contribution to noise is smoothed with a Gaussian kernel of $\sigma=10$\,\AA\ to imitate the DESI pipeline \cite{guySpectroscopicDataProcessingPipeline2022}.

Previously, in the DESI EDR analysis, the output resolution matrix was erroneously calculated using a box-car average of the finely sampled camera resolution matrix, both over rows and columns. This smoothed the resolution matrix twice and required a deconvolution as a fix in the estimator. We corrected this double-smoothing issue in the output resolution matrix, so that the deconvolution step described in refs.~\cite{karacayliOptimal1dDesiEdr2023, ravouxFFTP1dEDR2023} is no longer applied to these mocks. However, the resolution matrix still needs to be oversampled by a factor of two to improve the accuracy in the quadratic estimator formalism. 

\begin{figure}
    \centering
    \includegraphics[width=\linewidth]{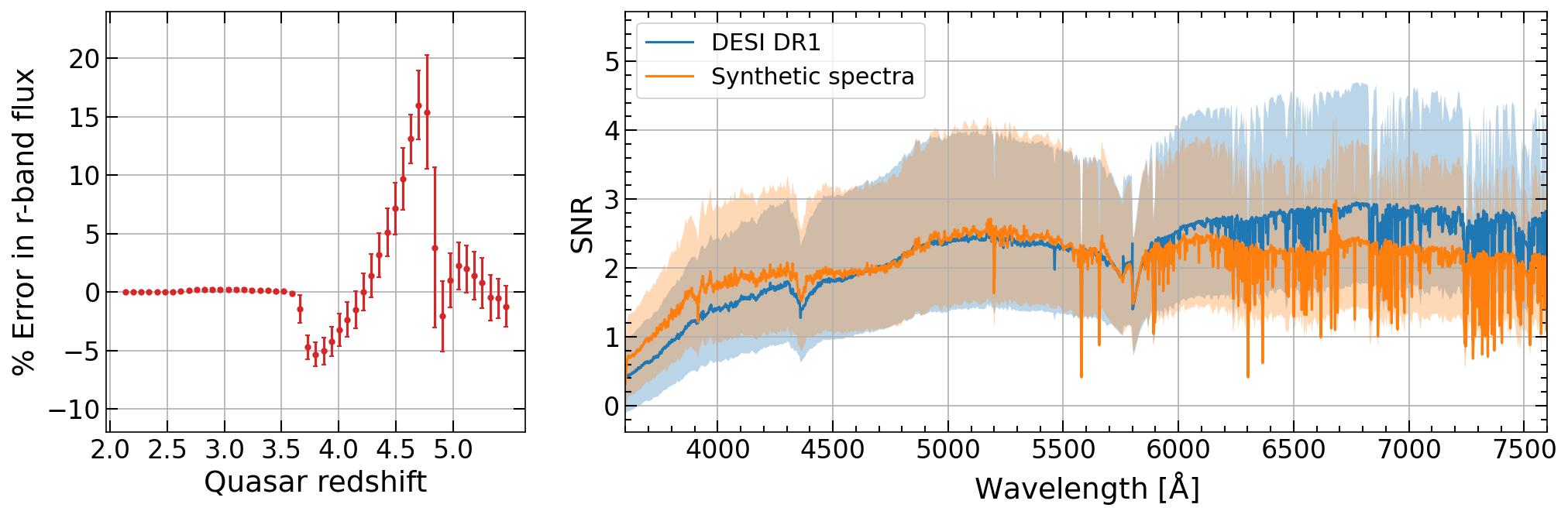}
    \caption{({\it Left}) The average percent error and scatter in the generated mock $r$-band flux values binned in redshift. We achieve a good match at $z_\mathrm{qso} < 4$, which constitutes the majority of the DR1 quasar sample. ({\it Right}) SNR with respect to observed wavelength. We achieve a good agreement between the data ({\it blue}) and the mocks ({\it orange}). Solid lines represent the median, and the area between the 32nd and 68th percentiles is shaded.}
    \label{fig:qq_fluxr_snr}
\end{figure}

\qq\ was further improved to match the noise properties of the data~\cite{herrera-alcantarDESIQuickquasars2023}. This was achieved by generating quasar continuum templates that match the $r$-band flux values and adding noise to each spectrum based on the effective exposure time $T_\mathrm{eff} = 12.15\times \mathrm{TSNR}^2_\mathrm{LRG}$~seconds, where $\mathrm{TSNR}_\mathrm{LRG}$ is the signal-to-noise ratio of the luminous red galaxy (LRG) template \cite{guySpectroscopicDataProcessingPipeline2022}. The input $r$-band flux values must correspond to the unabsorbed quasar continuum, although the \lya\ forest begins to affect the $r$-band for quasars at $z > 3.7$. We correct this effect by amplifying $r$-band flux values based on the average mean flux decrement caused by the forest. The relative error in the generated mock $r$-band flux values is shown in the left panel of figure~\ref{fig:qq_fluxr_snr}, where the data points represent the mean error in a redshift bin and error bars represent the scatter in that bin. We achieve a good match for the majority of our sample ($z_\mathrm{qso} < 4$) and a maximum of 15\% offset on average around $z=4.7$. The right panel of figure~\ref{fig:qq_fluxr_snr} demonstrates that the wavelength dependence of SNR in the synthetic spectra parallels the data. All unique targets identified as quasars are simulated in our mocks at their exact redshifts, approximately at the same $r$-band flux values and noise properties.

We randomly insert high-column density systems (HCDs), including damped \lya\ absorption systems (DLAs), without any correlations with the underlying matter field. The number of systems per unit redshift is set to be $l(z) = 0.96 e^{-7/z^2}$ following ref.~\cite{prochaskaSdssDr5Dlas2008}. We draw column densities following the column density distribution presented in ref.~\cite{prochaskaTowardsUnifiedIgm2014} between $\log N_\mathrm{H\textsc{i}} \in [19.5, 22.5]$. In reality, these systems are expected to cluster with the underlying density field \cite{mcdonaldPhysicalDampingWings2005, rogersEffectsOfHcd2018}. Therefore, the effect of clustering and incompleteness of real data catalogs still needs to be included in cosmological analyses \cite{palanque-delabrouilleOnedimensionalLyalphaForest2013, chabanierOnedimensionalPowerSpectrum2019}.

We insert broad absorption lines (BALs) into 16\% of quasars \cite{filbertBALEDRcatalog2023, martiniDesiBalY12024}. Unlike the validation tests for Baryon Acoustic Oscillations studies \cite{cuceuValidationofDesi2024LyaBao2024}, we do not introduce metal lines or higher-order Lyman lines.

\subsection{CCD image simulations}
\label{subsec:ccd_image_sims}

We produced CCD image simulations to validate the implementation of the resolution matrix by the spectroscopic pipeline \citep{guySpectroscopicDataProcessingPipeline2022}. This is critical as the accuracy of the resolution matrix is the dominant source of systematic error on small spatial scales. The simulations we produced for this work were constructed in the same manner as those described in ref.~\cite{karacayliOptimal1dDesiEdr2023} for the EDR analysis \cite{karacayliOptimal1dDesiEdr2023,ravouxFFTP1dEDR2023}, so here we only provide a short description of how we generate the spectra and then focus on the key differences from our previous work. 

We produced the image simulations with the \texttt{desisim} package, which was built by the DESI collaboration to develop and validate the spectroscopic data processing pipeline \citep{guySpectroscopicDataProcessingPipeline2022}. This package can produce realistic versions of the raw, two-dimensional CCD data that are the starting point for the pipeline processing. The inputs to this code are the intrinsic spectra of targets for a single DESI tile with apparent magnitudes representative of DESI quasar targets. The code models the bias, gain, and readnoise of the detectors, as well as the spectroscopic PSF, trace location, and throughput as a function of wavelength. The code also includes a realistic model of night sky emission and spectroscopic standards. 

As in ref.~\cite{karacayliOptimal1dDesiEdr2023}, we populate the tiles with only high-redshift quasars (4500 per tile), along with a selection of the standard stars and sky fibers that are required for the operation of the spectroscopic pipeline. As the DESI DR1 sample is substantially larger than EDR, we construct 150 simulated tiles of DESI quasars. Our final dataset is, therefore, 675,000 quasars, which is over an order of magnitude larger than the number we created for EDR and about 50\% more than the number of DESI DR1 quasars at $z > 2.09$. Another significant change relative to the EDR simulations is that we changed the redshift distribution from a uniform distribution over $2.6 < z < 3.6$ to one that is a better match to the observations. We also generated mock arc lamp spectra, which are used by the pipeline to measure the PSF and ultimately derive the resolution matrix. Finally, we processed the simulated dataset with the same pipeline we used for the observations.

\section{Methods\label{sec:methods}}

\subsection{Continuum fitting}
The power spectrum estimation requires the transmitted flux fluctuations $\delta_{F}$ due to \lya\ forest absorption. For a quasar $q$ at redshift $z_q$, this is given by 
\begin{equation}
    \delta_{F, q}(\lambda) = \frac{f_q(\lambda)}{\overline{F}(z_\alpha)C_q(\lambda_\mathrm{RF})} - 1,
\end{equation}
where $\lambda$ is the observed wavelength, $f_q(\lambda)$ is the observed flux, and $\lambda_\mathrm{RF} = \lambda / (1+z_q)$ is the wavelength in the quasar's rest frame. Since  $\delta_{F, q}$ is relative to the mean transmission of the IGM ($\overline{F}$), the quasar continuum $C_q(\lambda_\mathrm{RF})$ is multiplied by $\overline{F}(z_\alpha)$ at the \lya\ forest redshift $z_\alpha = \lambda / \lambda_\mathrm{Ly\alpha} -1$. 

In real data, the quasar continuum is unknown and needs to be estimated. We use the same continuum fitting algorithm that was developed and has been applied to both $\xi_\mathrm{3D}$ and \poned\ measurements to compute $\delta_{F, q}$ \cite{bourbouxCompletedSDSSIVExtended2020, karacayliOptimal1dDesiEdr2023}. In this formalism, the definition of the quasar continuum absorbs the mean transmission of the IGM, such that the quasar ``continuum" $\overline{F}(z_\alpha) C_q(\lambda_\mathrm{RF})$ is given by
\begin{equation}
    \overline{F}(z_\alpha)C_q(\lambda_\mathrm{RF}) = \overline{C}(\lambda_\mathrm{RF}) \left( a_q + b_q \Lambda \right)\, , \quad \mathrm{where}\quad
    \Lambda \equiv \frac{\log\lambda_\mathrm{RF} - \log\lambda_\mathrm{RF}^{(1)}}{\log\lambda_\mathrm{RF}^{(2)} - \log\lambda_\mathrm{RF}^{(1)}},
\end{equation}
$\lambda_\mathrm{RF}^{(1, 2)}$ are the minimum and maximum wavelengths considered in the calculation, $\overline{C}(\lambda_\mathrm{RF})$ is the global mean continuum, and $a_q$ and $b_q$ are two quasar diversity parameters. These quasar diversity parameters also absorb the mean IGM transmission.

The overview of our method is as follows (see refs.~\cite{bourbouxCompletedSDSSIVExtended2020, ramirezperezLyaCatalogDesiEdr2023, karacayliOptimal1dDesiEdr2023} for a detailed description). For each quasar, we find the $a_q$ and $b_q$ values that minimize the following cost function while keeping \textit{all} other parameters fixed:
\begin{equation}
    \chi^2 = \sum_j \frac{\left[f_j - ( a_q + b_q \Lambda_j) \overline{C}\left(\frac{\lambda_j}{1 + z_q}\right) \right]^2}{\sigma_{q, j}^2} + \sum_j \ln \sigma_{q, j}^2,
\end{equation}
where the summation $j$ is over all pixels in the forest region and $\lambda_j$ is the observed wavelength.
The major complication comes from $\sigma_{q, j}^2$, which must take into account the intrinsic large-scale \lya\ fluctuations $\sigma^2_\mathrm{LSS}$:
\begin{equation}
    \sigma_{q, j}^2 = \eta(\lambda_j) \sigma^2_\mathrm{pipe, j} + \sigma^2_\mathrm{LSS}(\lambda_j) ( a_q + b_q \Lambda_j)^2 \overline{C}^2\left(\frac{\lambda_j}{1 + z_q}\right) \label{eq:sigma2_cfit},
\end{equation}
where $\eta(\lambda)$ is the pipeline noise correction term. The FFT estimator uses a specific case in which $\sigma_{q, j}$ is assumed constant across the quasar spectrum. 
After every quasar is fit, we stack all continua in the rest frame, update the global mean continuum  $\overline{C}$, and calculate $\eta$ and $\sigma^2_\mathrm{LSS}$ as described in the references. This algorithm is implemented in two separate packages: \texttt{picca}\footnote{\url{https://github.com/igmhub/picca}}~\cite{bourbouxCompletedSDSSIVExtended2020} and \texttt{qsonic}\footnote{\label{footnote:picca}\url{https://qsonic.readthedocs.io/en/stable/}}~\cite{Karacayli_QSOnic_fast_quasar_2024}. The FFT estimator uses \texttt{picca} as it has been part of the same package, whereas the optimal estimator uses \texttt{qsonic} since it is significantly faster. We also utilize one of \texttt{qsonic}'s differentiating features---employing a fiducial mean IGM transmission to mitigate its coupling to quasar diversity parameters.

\subsection{Optimal estimator}
The quadratic maximum likelihood estimator (QMLE) formalism has been a highly successful technique in power spectrum measurements from the cosmic microwave background radiation, galaxy surveys, and weak lensing \cite{hamiltonOptimalMeasurementPower1997, tegmarkKarhunenLoeveEigenvalueProblems1997, tegmarkMeasuringGalaxyPower1998, seljakWeakLensingReconstruction1998}. It was also successfully applied to measure \lya\ forest \poned\ from a variety of datasets such as 3,000 quasars from SDSS~\cite{mcdonaldLyUpalphaForest2006}, 527 high-resolution, high-SNR spectra~\cite{karacayliOptimal1DLy2022}, and 54,600 quasars from the DESI early data release~\cite{karacayliOptimal1dDesiEdr2023}. We call this the optimal estimator.

The optimal estimator works in real space (instead of Fourier space) to estimate the power spectrum and therefore allows weighting by the pipeline noise, accounts for intrinsic \lya\ large-scale structure correlations, and, most importantly, is not biased by masking unwanted pixels in the spectra.
We refer the reader to refs.~\cite{karacayliOptimal1DLy2020, karacayliOptimal1DLy2022} for more details about our development process and application to high-resolution spectra. In short, the optimal estimator gives a power spectrum $\bm{p}_q$ and its covariance matrix $\mathbf{C}_q$ for each quasar $q$. This covariance matrix $\mathbf{C}_q$ includes contributions from the pipeline noise estimates, the signal contribution based on a fiducial power spectrum, and the ``survey window function" due to masking and continuum marginalization. The final \poned\ estimate is a weighted average: $\bm{p} = \mathbf{C}_\mathrm{tot} \sum_q \mathbf{C}^{-1}_q \bm{p}_q$, where $\mathbf{C}_\mathrm{tot}^{-1} \equiv \sum_q \mathbf{C}^{-1}_q$. The inverse covariance matrix is also referred to as the Fisher matrix $\mathbf{F} = \mathbf{C}^{-1}$, and as the mode-mixing matrix in the CMB literature. This matrix normalizes the power spectrum in the weighted-average sense and also deconvolves the survey window function, which mixes different $(k, z)$ modes due to masking, continuum marginalization, and the reverse interpolation of pixel pairs into two redshift bins.

We have made large performance improvements to our code since DESI's early data analysis. These updates include: 1) Using a tile strategy while transpose-copying a matrix, which is ten times faster than the straightforward implementation. 2) Calculating $\mathbf{Q}_\alpha=\mathbf{C}^{-1}\mathbf{C}_{,\alpha}$ instead of $\mathbf{C}^{-1} \mathbf{C}_{,\alpha} \mathbf{C}^{-1}$, which reduces the number matrix multiplications. However, the trace calculation for the Fisher matrix $F_{\alpha\beta}=0.5~\mathrm{Tr} (\mathbf{Q}_\alpha \mathbf{Q}_\beta^\mathrm{T})$ can only be efficiently performed using the fast transpose update of point 1, leveraging the dot product. 3) Fine-tuning the shared-memory parallelization with \texttt{OpenMP}. Thanks to these performance gains, the estimator is now four times faster, and we are now able to estimate \poned\ at 85 $k$ bins per redshift bin and perform repeated runs on many mock realizations of DR1.

\subsubsection{Estimating the covariance matrix\label{subsec:qmle_cov}}
Another major numerical update concerns the bootstrap covariance matrix. Our code was limited to using each MPI task as a subsample in DESI early data analysis. This largely yielded noisy covariance matrix estimates since a few hundred independent samples cannot fully explore all the fluctuations in an approximately $ 1000\times1000$ matrix. We had to resort to strong sparsity cuts to suppress this noise, which prevented exploring correlated systematics fluctuations. We now bootstrap over quasars to fully examine the covariance matrix. However, we still have to make a few simplifying assumptions to speed up this procedure. The main approximation concerns the Fisher matrix. Calculating this matrix for every bootstrap realization is computationally expensive. Instead, we use the same Fisher matrix for all realizations. Second, we assign an observation frequency for each quasar based on a Poisson distribution with a mean of one \citep{hanleyNonParametricPoissonBootstrap2006, chamandy2012estimating}. This means we do not have the same number of quasars across bootstrap realizations, but ref.~\cite{hanleyNonParametricPoissonBootstrap2006} has shown that this introduces a negligible bias for a large number of independent data points ($n \gtrsim 100$). Finally, we can directly estimate the inverse covariance matrix (i.e. the Fisher matrix), which is what likelihood inferences need, using the constant Fisher matrix approximation and the fact that under the assumption of Gaussianity,  $\langle \bm{y} \bm{y}^\mathrm{T}\rangle = \mathbf{C}^{-1}$ where $\bm{y} = \mathbf{C}^{-1} \bm{p}$. In other words, we calculate the covariance of the weighted data vectors between bootstrap realizations, which is approximately equal to the inverse of the covariance matrix of \poned.

Unfortunately, the resulting bootstrap is still susceptible to noise problems, mostly at higher redshifts where there are only a few hundred quasars. However, we no longer need strong sparsity cuts thanks to our new regularization scheme that can reliably preserve the fluctuations due to systematics while suppressing the noise in the bootstrap inverse covariance matrix estimates $\mathbf{F}_\mathrm{B}$. The optimal estimator formalism calculates the Fisher matrix under the assumption of Gaussianity, which we will refer to as $\mathbf{F}_\mathrm{G}$ below. We compute the final bootstrap covariance as follows:
\begin{enumerate}
    \item Using the diagonals of the Gaussian matrix $v_i = \sqrt{F^\mathrm{G}_{ii}}$, normalize both matrices:
    \begin{equation}
        R^\mathrm{G}_{ij} = F^\mathrm{G}_{ij} / v_i v_j \,, \qquad R^\mathrm{B}_{ij} = F^\mathrm{B}_{ij} / v_i v_j \, .
    \end{equation}

    \item Subtract the two correlation matrices $\Delta \mathbf{R} = \mathbf{R}_\mathrm{B} - \mathbf{R}_\mathrm{G}$ and smooth it using a 2D Gaussian filter with $\sigma=2$. The first half of the redshift bins is smoothed box by box to eliminate edge effects. This step yields a smooth correlation difference matrix $\widetilde{\Delta \mathbf{R}}$.

    \item Reconstruct the smoothed bootstrap matrix $\widetilde{\mathbf{F}}_\mathrm{B}$:
    \begin{equation}
        \widetilde{R}^\mathrm{B}_{ij} = R^\mathrm{G}_{ij} + \widetilde{\Delta R}_{ij}\, \quad \rightarrow \quad\, \widetilde{F}^\mathrm{B}_{ij} = v_i v_j \widetilde{R}^\mathrm{B}_{ij}\, .
    \end{equation}
    \setcounter{enumlast}{\value{enumi}}
\end{enumerate}

Theoretically, the Gaussian covariance matrix must yield the smallest errors. In more formal terms, the inequality $\bm{x}^\mathrm{T} \mathbf{C}_\mathrm{B} \bm{x} \geq \bm{x}^\mathrm{T} \mathbf{C}_\mathrm{G} \bm{x}$ must hold for all vectors $\bm{x}$. The smoothing procedure described so far cannot guarantee this theoretical boundary by itself. We now force this lower limit in two more steps.

\begin{enumerate}
    \setcounter{enumi}{\value{enumlast}}
    \item We find the eigenvalues $\lambda_i$ and eigenvectors $\bm e_i$ of this smoothed bootstrap inverse covariance matrix $\widetilde{\mathbf{F}}_\mathrm{B}$. We calculate
    the values corresponding to each eigenvector using the Gaussian matrix $\lambda_i^{\mathrm{G}} = \bm e_i^T \mathbf{F}_{\mathrm{G}} \bm e_i$. These are the theoretical maxima, therefore we replace $\lambda_i \rightarrow \mathrm{min}(\lambda_i, \lambda_i^{\mathrm{G}})$ and reconstruct $\widetilde{\mathbf{F}}_\mathrm{B}$ with these new eigenvalues. This procedure is the same as refs.~\cite{mcdonaldLyUpalphaForest2006, karacayliOptimal1dDesiEdr2023}.

    \item The inequality written above means that $\mathbf{M} = \widetilde{\mathbf{C}}_\mathrm{B} - \mathbf{C}_\mathrm{G}$ must be a positive semi-definite matrix. We calculate $\mathbf{M}_+$ by setting all negative eigenvalues of $\mathbf{M}$ to zero, and obtain the final bootstrap covariance matrix: $\widetilde{\mathbf{C}}_\mathrm{B}^\mathrm{final} = \mathbf{C}_\mathrm{G} + \mathbf{M}_+$. Note that this step cannot be done with inverse covariance matrices.
\end{enumerate}

\begin{figure}
    \centering
    \includegraphics[width=\columnwidth]{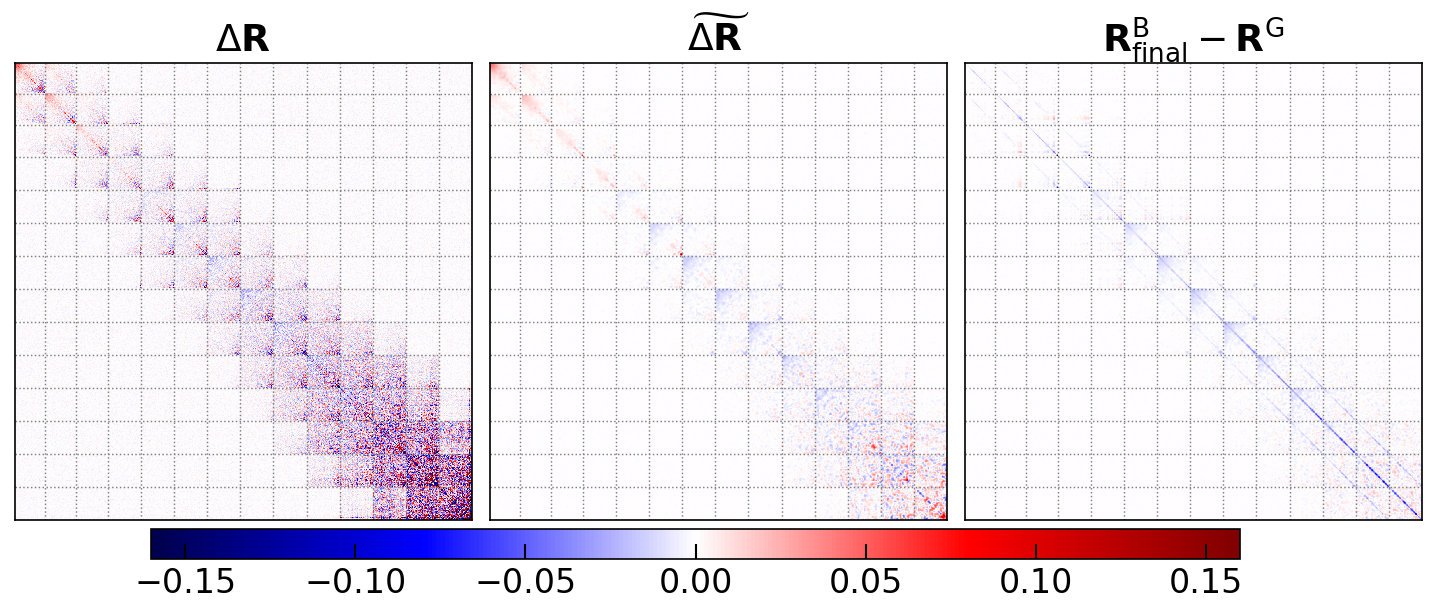}
    \caption{The difference matrices $\Delta \mathbf{R}$ at three stages of the optimal estimator's regularization scheme. The redshift increases from left (top) to right (bottom) in each panel. The initial estimate shown in the left panel has strong off-diagonal terms in the first few redshift bins and has significant noise at higher redshifts. The middle panel shows the 2D smoothed difference matrix $\widetilde{\Delta \mathbf{R}}$ in which the noise is largely suppressed. The right panel shows the difference for the final bootstrap estimate after it is forced to be larger than the Gaussian covariance matrix.}
    \label{fig:qmle_boot_steps}
\end{figure}
The left panel in figure~\ref{fig:qmle_boot_steps} shows the initial difference matrix $\Delta \mathbf{R}$, where the redshift increases from left (top) to right (bottom). The first two redshift bins exhibit strong off-diagonal contributions in the bootstrap estimate, whereas higher redshift bins are significantly noisier. The middle panel shows the 2D smoothed difference matrix $\widetilde{\Delta \mathbf{R}}$. The noise is largely suppressed. To preserve the off-diagonal structure of the first few redshift bins at the edges, we smooth each redshift bin separately. The right panel shows the difference for the final bootstrap estimate after forcing the inequality $\bm{x}^\mathrm{T} \mathbf{C}_\mathrm{B} \bm{x} \geq \bm{x}^\mathrm{T} \mathbf{C}_\mathrm{G} \bm{x}$.

\subsubsection{Analysis choices}
We calculate the transmitted flux fluctuations $\delta_F$ using \texttt{qsonic} with the following settings. We set the rest-frame wavelength range to $1050-1180~$\AA\ following the literature standard, conservatively avoiding the large, hard-to-model profiles of the \lya\ and Ly$\beta$ emission lines. We limit the observed wavelength range to $3600-7000~$\AA\ based on the minimum wavelength measured and the most distant observed quasar redshift ($z_\mathrm{qso}\approx 5$) by DESI. The global mean continuum $\overline{C}(\lambda_\mathrm{RF})$ is measured in 0.4~\AA\ steps. We select quasars with an average SNR greater than one at wavelengths greater than the \lya\ emission line of the quasar and with an average SNR greater than 0.3 in the forest region. We do not apply any pipeline noise correction, but we still fit for $\eta$ during continuum fitting.

When estimating \poned\, we split spectra into two segments if they have more than 350 pixels, and we ignore segments with fewer than 20 pixels to reduce computation time and improve continuum marginalization. We use 60 linear bins with $\Delta k_\mathrm{lin}=5\times10^{-4}~$\skm, which is comparable to the fundamental frequency of one segment, followed by 25 log-linear bins with $\Delta k_\mathrm{log}=0.01$, reaching a maximum $k$ that remains smaller than the Nyquist frequency. We measure \poned\ with redshift bins of size $\Delta z=0.2$ from $z=2.0$ to $z=4.6$. As noted in the DESI EDR analysis in ref.~\cite{karacayliOptimal1dDesiEdr2023}, the $z=2.0$ bin is susceptible to noise and interpolation errors, so we exclude it from all analyses presented in this work. We use the true power spectrum as a fiducial to calculate the signal matrix~\cite{karacayliOptimal1DLy2020}, and interpolate the signal and derivative matrices using sufficiently high resolution in redshift and velocity spacings.

\subsubsection{Smoothing the measured power spectrum}
The corrections proposed in this paper for the optimal estimator are scalings of a smooth power spectrum $P_{\mathrm{smooth}}$ $(k, z)$ following ref.~\cite{karacayliOptimal1dDesiEdr2023}. This $P_{\mathrm{smooth}}$ is calculated
by applying \texttt{SmoothBivariateSpline} to the logarithms of $1+z, k$ and measured (noisy) \poned\ with corresponding statistical weights $\sigma_\mathrm{boot}/$\poned\ and a smoothing factor that is five times the number of data points.
This function in \texttt{scipy} is based on the algorithm presented in refs.~\cite{DierckxAlgorithmforSurface1981, DierckxCurveandSurface1993}.

\subsection{FFT estimator}

The FFT estimator was used in several studies to directly compute \pk~from a set of transmitted flux fluctuations~\cite{mcdonaldPredictingLyaPower2003,mcdonaldLyUpalphaForest2006,palanque-delabrouilleOnedimensionalLyalphaForest2013,chabanierOnedimensionalPowerSpectrum2019,ravouxFFTP1dEDR2023}. The complete description of the FFT estimator used in this study is given in ref.~\cite{ravouxFFTP1dEDR2023}. We provide a brief description and present the modifications below. The core concept of the FFT estimator is to decompose $\delta_{F,q}(\lambda)$ for each quasar $q$ in the following manner: 

\begin{equation}
    \label{eq:contrast_p1d}
    \delta_{F,q} (\lambda)=  \left(\delta_{\mathrm{Ly\alpha},q} (\lambda)+\delta_{\mathrm{metals},q} (\lambda)\right)\circledast W_q(\lambda,\mathbf{R}) + \delta_{\mathrm{noise},q}(\lambda)\,,
\end{equation}
where $\delta_{\mathrm{\lya},q}$ is the \lya~contrast we want to measure, $\delta_{\mathrm{metals},q}$ corresponds to the contrast caused by metal absorption in the \lya~forest region, $\delta_{\mathrm{noise},q}$ is the noise contrast, and $W_q(\lambda,\mathbf{R})$ is the effect of the finite resolution of the spectrograph on both astrophysical signals. It is derived from the resolution matrix $\mathbf{R}_{q}$, which characterizes the spectrograph resolution \cite{boltonSpectroPerfectionismAlgorithmicFramework2010, guySpectroscopicDataProcessingPipeline2022}. Crucially, this estimator is employed in the wavelength configuration, as DESI processes the spectra in an equally spaced wavelength grid.

Following ref.~\cite{ravouxFFTP1dEDR2023}, the derived \lya~contrast is cut into segments of the same length, noted as $s$. We Fourier transform this contrast and account for the different correlations in Fourier space between the signals mentioned above. The FFT estimator for the redshift bin $z$ and the wavenumber bin $A$ is as follows: 

\begin{equation}
\label{eq:fft_p1d_estimator}
P_{\mathrm{1D}}(A,z) = \left\langle\left[P_{\mathrm{raw},s}(k)-P_{\mathrm{noise},s}(k) \right] \cdot \mathbf{R}_s^{-2}(k) \right\rangle_{s \in z, k\in A} - P_{\mathrm{metals}}(A) \,,
\end{equation}

\noindent where the ensemble average is calculated over all segments and wavenumbers that fall into the bin $(z, A)$, and $P_{\mathrm{raw},s}$ is the direct Fourier transform of the transmitted flux fluctuation in the \lya~forest region, $P_{\mathrm{noise},s}$ is the noise power spectrum estimated directly from the pipeline, and $\mathbf{R}_s$ is the Fourier transform of the resolution matrix for the segment $s$. Note that the wavenumbers are naturally in \AA$^{-1}$ units, while the power spectra are in \AA\ units.
Since our mocks have no metal contamination, the $P_{\mathrm{metals}}$ term is ignored in this work.  Finally, our objective is to validate and evaluate the astrophysical and instrumental systematics correction terms in the final measurement of \pk\ presented in ref.~\cite{ravouxFFTP1dDesiDr12024}.

Before computing the FFT estimator, we apply an average signal-to-noise ratio cut $\overline{\mathrm{SNR}} > 1$ per pixel, where the $\overline{\mathrm{SNR}}$ is defined as the average over the considered part of the \lya~forest. To get thinner redshift bins without too much correlation, the $\delta_{F}$'s are split into three consecutive and non-overlapping segments of equal length. We remove segments shorter than 75 spectrum pixels due to a cut in the UV region or the presence of a DLA. We also do not consider the \lya\ segments with more than 120 masked spectrum pixels. We compute the FFT \pk~estimator by averaging (as shown with the $\langle\cdot\rangle$ operation in eq.~\eqref{eq:fft_p1d_estimator}) over individual \lya\ segments. For convenience, we often refer to the term inside the averaging operator as an ``individual power spectrum", denoted $p_{s}(k)$ for the segment $s$ and the wavenumber $k$, although this is not mathematically a power spectrum.

When averaging individual power spectra within a given redshift and wavenumber bin, we apply weights $w_{A, s} = 1/\sigma^2 (A, s)$ based on each segment's SNR as described in ref.~\cite{ravouxFFTP1dEDR2023}.  Here, $\sigma^2 (A, s)$ represents the variance associated with each segment $s$ and wavenumber bin $A$, and is fitted as a function of the segment’s SNR.  The eq.~\eqref{eq:fft_p1d_estimator} is then calculated as a weighted average over all individual power spectra's $k$ wavenumbers that contribute to the wavenumber bin $A$:

\begin{equation}
\label{FFT-mean}
\left\langle P \right \rangle(A,z) = \frac{\sum_{s \in z, k\in A} w_{A,s} p_s (k) }{\sum_{s \in z, k\in A} w_{A,s} }\, .
\end{equation}

Contrary to ref.~\cite{ravouxFFTP1dEDR2023}, we do not use the error bars obtained from the $\mathrm{SNR}$ weighting scheme but choose to estimate the statistical error bars with the following unbiased estimator:

\begin{equation}
\label{FFT-var}
\langle \sigma\rangle^2 (A,z) =  \left(\frac{(\sum_{s \in z} w'_{A,s})^2 }{\sum_{s \in z} (w'_{A,s})^2}  -1\right)^{-1} \left[ \frac{\sum_{s \in z} (w'_{A,s})^2 P_{A,s}^2}{\sum_{s \in z} (w'_{A,s})^2} - \left\langle P_{A,s} \right\rangle_{s \in z}^2 \right]\,,
\end{equation}

\noindent where $P_{A,s} = \sum_{k \in A_s} p_s(k) / N_{A,s}$, $A_s$ is the ensemble of wavenumbers associated with segment $s$ that fall into bin $A$,  $N_{A,s}$ is the number of wavenumbers in $A_s$, and $P_{A,s}$ is the average of individual power spectra in bin $A$. The weights used here are defined as $w_{A,s}^\prime = N_{A,s} w_{A,s}$.

\subsubsection{Estimating the covariance matrix}

We develop a new, unbiased estimator for the full covariance matrix that accounts for the aforementioned SNR weighting scheme:

\begin{align}
\label{FFT-cov}
\mathcal{\langle C \rangle}(A, B,z) = & \left(\frac{\sum_{s \in z} w'_{A,s} \sum_{s \in z} w'_{B,s}}{\sum_{s \in z} w'_{A,s} w'_{B,s}}  -1\right)^{-1} \nonumber \\
&\left[ \frac{\sum_{s \in z} w'_{A,s} w'_{B,s} P_{A,s} P_{B,s}}{\sum_{s \in z} w'_{A,s} w'_{B,s}} - \left\langle P_{A,s} \right\rangle_{s \in z} \left\langle P_{B,s} \right\rangle_{s \in z}\right]\,.
\end{align}

After ascribing a redshift bin to each segment, the averaging of individual power spectra, along with the computation of the variance and the covariance, is performed independently for each redshift bin.

\subsection{Kolmogorov–Smirnov test\label{subsec:kstest}}
Our tests on mocks are based on calculating $\chi^2_i$ for each realization $i$:
\begin{equation}
\label{eq:chi2}
\chi^2_i = (\bm p_i - \bm p_{\mathrm{ref}})^\mathrm{T} \mathbf{C}^{-1}_i (\bm p_i - \bm p_{\mathrm{ref}}),
\end{equation}
where $\mathbf{C}_i$ is the estimated covariance matrix for realization $i$, and $p_\mathrm{ref}$ is the reference power spectrum. The latter is the analytically calculated power spectrum for the optimal estimator, as both are naturally in velocity configuration \cite{karacayliOptimal1DLy2020}. However, the FFT estimator is constructed in wavelength configuration, and the conversion from velocity to wavelength configuration introduces slight biases. So, $p_\mathrm{ref}$ instead needs to be measured from mocks for the FFT measurement.

These $\chi^2$ values are inherently random, drawn from an ``empirical" distribution with an underlying cumulative distribution function (CDF): $F(\chi^2)$. We will compare this to a theoretical distribution using the Kolmogorov-Smirnov (KS) test. The KS test is a nonparametric test that does not require any assumptions about $F(\chi^2)$ when comparing it against other probability distributions. It can reject the chosen null hypothesis if the p-value is below a specified threshold. Typically, this threshold is set to 0.05, balancing the trade-off between false negatives and false positives. We show below that this remains a reasonable threshold for detecting a few percent biases with $95\%$ efficiency.


Our theoretical distribution is a $\chi^2$ probability distribution with the given number of degrees of freedom of the data vector, $\nu$. Let us denote the CDF of this distribution as $G(\chi^2)$. We adopt the one-sided KS test with the following:
\begin{itemize}
    \item \textbf{Null hypothesis:} Results are unbiased, and errors are correctly estimated or overestimated, $F(\chi^2)\geq G(\chi^2)$.
    \item \textbf{Alternative hypothesis:} Results are biased, or errors are underestimated, $F(\chi^2) < G(\chi^2)$ for some $\chi^2$.
\end{itemize}
We reject the null hypothesis (therefore fail the validation test) if the KS test yields a p-value less than 0.05. This is similar to goodness-of-fit tests, where higher p-values indicate a better fit. Due to computational time limitations in analysis, we generate \numMocks\ independent realizations, which limits the power of the KS test.

We first note that our p-value threshold of 0.05 sets an upper bound of 95\% on the true-positive rate, even when the null hypothesis holds (p-values are uniformly distributed between zero and one in this case). We run numerical trials to further understand how effective this choice of p-value is against a possible global bias factor on covariance. We run 10,000 numerical trials, where we generate \numMocks\ random numbers from the $\chi^2(\nu=1035)$ distribution in each trial. We find that, with these choices, we can rule out a 2.85\% bias in 80\% of trials and a 3.75\% bias in 95\% of trials. In other words, there remains at least $5\%$ chance to falsely pass the validation test even though the distribution was biased by a few percent.
A smaller p-value of $0.01$ would reduce the chance of falsely rejecting the null hypothesis but would require a bias as large as 3.69\% to rule it out in 80\% of the trials. In contrast, a larger p-value of $0.1$ can rule out a 2.41\% bias in 80\% of trials, but at the cost of a higher chance of falsely rejecting (failing) the null hypothesis (validation test). The most effective way to improve the KS test is to increase the number of mocks, which we leave to future work.

Lastly, we occasionally quote the two-sided KS test results to understand whether errors are overestimated. The null hypothesis in this case is $F(\chi^2)= G(\chi^2)$, i.e., the errors are correctly estimated. We can logically conclude that the errors are overestimated if the one-sided test passes and the two-sided test rejects the null hypothesis.


\section{Validating the optimal estimator\label{sec:qmle_val}}
We start our analysis by validating the optimal estimator. Because of its insensitivity to masking and consequently the survey window function, the list of tests simplifies to tests of uncontaminated data and fully contaminated data.

Even though the optimal estimator can marginalize over the dominant continuum fitting errors, it cannot completely eliminate them. These errors add additional power to all $k$ scales through higher-order correlations. We derive this bias for the optimal estimator and examine the effect of masked pixels on the induced bias.

\subsection{Validation in the absence of contaminants\label{subsec:qmle_val_nosyst}}

This section aims to (1) validate the estimators when no contaminants are present and the quasar continuum is perfectly known, (2) validate the covariance matrix, and (3) derive the biases resulting from the continuum fitting procedure. We generate and use \numMocks\ \qq\ mocks without any contaminants for this goal.

\begin{figure}
    \centering
    \includegraphics[width=0.37\columnwidth]{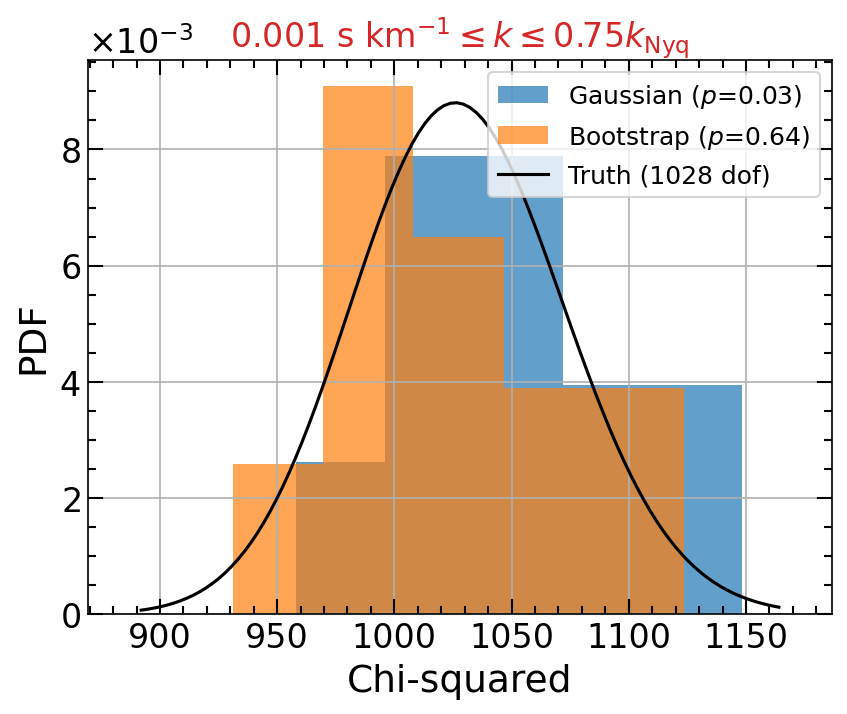}
    \includegraphics[width=0.62\columnwidth]{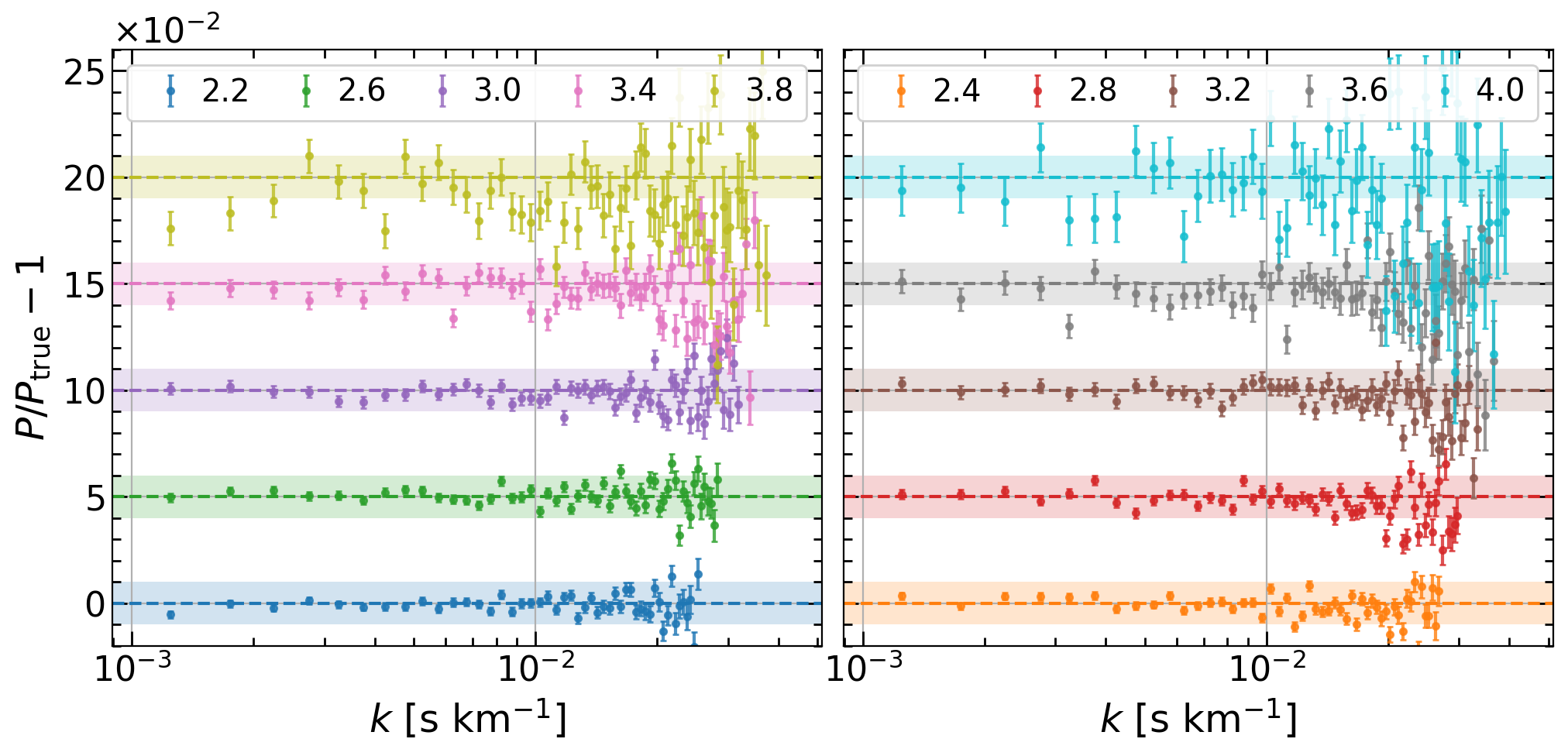}
    \caption{({\it Left}) PDF of $\chi^2-$squared values for the no-systematics case from the optimal estimator. The Gaussian covariance matrix fails the KS test by yielding a p-value of 0.03, while the bootstrap covariance matrix passes it by yielding a p-value of 0.64. ({\it Right}) The ratio between the estimated \poned\ from the stack of \numMocks\ mock realizations and the true power minus one. The ratios are shifted vertically for clarity. Dashed lines indicate the expected value of zero for each redshift bin. The area within 1\% of the expected value is shaded. The fluctuations are between $ 0.5\%$ and $1\%$ for $z\leq 3.6$. Higher redshifts fluctuate more, so they are not shown for clarity.}
    \label{fig:pdf20_nosyst_qmle}
\end{figure}

We use the underlying true continuum of quasars and mean flux to derive transmitted flux fluctuations, which bypasses the continuum fitting step. The purpose of this test is to confirm that our pipeline does not introduce any small numerical errors due to factors such as interpolation, weighting, and random noise. We run the quadratic estimator without marginalization, as no continuum fitting errors exist. However, in the case of real data with continuum marginalization, $k<0.001~$\skm\ bins will become unusable due to continuum fitting, as we will discuss shortly. Additionally, high-$k$ bins will be limited by the spectrograph resolution, which we quantify with respect to effective velocity spacing $R_z=c\Delta\lambda_\mathrm{DESI}/(1+z)\lambda_\mathrm{\lya}$ of each redshift bin, where $\Delta\lambda_\mathrm{DESI}=0.8~$\AA. So, we limit the $k$ modes to the following range: $0.001~$\skm$\leq k \leq 0.75~ k_\mathrm{Nyq}(z)$, where the Nyquist frequency is defined as $k_\mathrm{Nyq}(z)=\pi/R_z$ and increases with redshift. We perform the KS test described in section~\ref{sec:methods}, and obtain p-values of 0.03 for the Gaussian covariance matrix and 0.64 for the bootstrap covariance matrix (as described in section~\ref{subsec:qmle_cov}), which is shown in figure~\ref{fig:pdf20_nosyst_qmle}. This indicates that the Gaussian covariance matrix does not fully capture the errors. Since the bootstrap matrix's p-value is greater than our 0.05 threshold, we consider this test passed by the optimal estimator.

We performed two additional checks. First, a two-sided KS test yields $p=0.88$ for the bootstrap covariance matrix, which indicates that it is likely a good description of the true covariance matrix.
Second, we stack the results from all \numMocks\ mock realizations by averaging \poned\ estimates. The resulting, more precise \poned\ follows the underlying true power, which is shown in the right panel of figure~\ref{fig:pdf20_nosyst_qmle}. This ratio reveals hints of subpercent-level biases in $3.8\leq z\leq 4.4$, and a 5\% bias at $z=4.6$ due to edge effects, which is not shown in the figure for clarity. Correcting for these biases or inflating the covariance matrices to account for these errors yields p-values greater than 0.9 for both cases, but likely overcorrects or overestimates. We will revisit this point when we include more realistic features in the mock spectra below.

The second goal is to investigate the effect of continuum fitting. We fit each quasar with $a_q + b_q\Lambda$ polynomial and marginalize out polynomials of $\Lambda$ from $\delta_F$ up to and including the second order. Besides contaminating low-$k$ modes, the continuum fitting adds extra power to all scales through three-point correlations. This bias is strongly detected with a p-value of approximately zero. Therefore, we need to quantify and correct this bias in our measurements. We model this as a relation with respect to the underlying true power $P_\mathrm{true}$ such that $P_\mathrm{meas}/P_\mathrm{true} - 1 = b(k)$, where $P_\mathrm{meas}$ is the measured \poned. To eliminate the statistical noise and possible systematic artifacts, we use the estimated \poned\ with true continuum in the denominator. We fit $b(k)$ from the stack of \numMocks\ mocks with a first-order polynomial of $k/(0.009~$\skm)\ between $0.001~$\skm$\leq k \leq 0.75~ k_\mathrm{Nyq}$ at each redshift bin. The resulting bias is $b(k) \sim 6 \times 10^{-3}$. Section~\ref{subsec:qmle_val_hcd_bal} provides more details about the robustness of these bias estimates.

We correct each power spectrum estimate using the smooth power spectrum as $P_\mathrm{corr} = P_\mathrm{meas} - \frac{b(k)}{1 + b(k)} P_{\mathrm{smooth}}$ for each redshift bin. After this correction, the bootstrap covariance matrix passes the KS test with a p-value of 0.1, whereas the Gaussian covariance matrix still fails. We conclude the optimal estimator results with bias correction, and the bootstrap covariance matrix is validated in the uncontaminated case. Since these mocks do not include systematics realistically, this correction is not applied to the data. The correction that is applied to the data is derived in the next subsection.

\subsection{Validation of masking HCDs, BALs, and atmospheric lines\label{subsec:qmle_val_hcd_bal}}

The real data is contaminated by BAL features, HCDs, and atmospheric emission lines, and is complicated by the ``survey window function" resulting from masking these features. The FFT estimator must correct for this window function (masking bias), whereas the quadratic estimator must confirm its robustness by construction. We generate \numMocks\ \qq\ mock realizations with BAL features and randomly placed DLAs to focus on the effect of masking specifically.

The optimal estimator method has two goals. The first goal is to confirm that masking these features does not introduce any biases. We calculate the transmitted flux fluctuations using the true continuum of each quasar to separate the effect of continuum fitting and mask all BAL features and HCDs at their exact locations.\footnote{As opposed to searching for their locations using a finder algorithm, which may produce incomplete, impure, and incorrect size estimates.} We further mask the regions around sky emission lines that will be masked in real data analysis. Then, we estimate \poned\ without any marginalization. We obtain p-values greater than 0.7 for both the Gaussian and bootstrap covariance matrices; therefore, the optimal estimator passes this test. The Gaussian matrix passing this test after previously failing may seem unexpected. However, this is due to weakening statistical power, as some pixels are ``thrown out", which results in diagonals of the Gaussian covariance matrix to be 30\% larger on average.

The second goal is to investigate the effect of continuum fitting when these features are masked. We follow the same steps outlined in section~\ref{subsec:qmle_val_nosyst}. The p-values are again zero without any correction. After the bias correction, the bootstrap covariance matrix passes the KS test with a p-value of 0.83, whereas the Gaussian covariance matrix fails with a p-value of 0.03.

\begin{figure}
    \centering
    \includegraphics[width=\columnwidth]{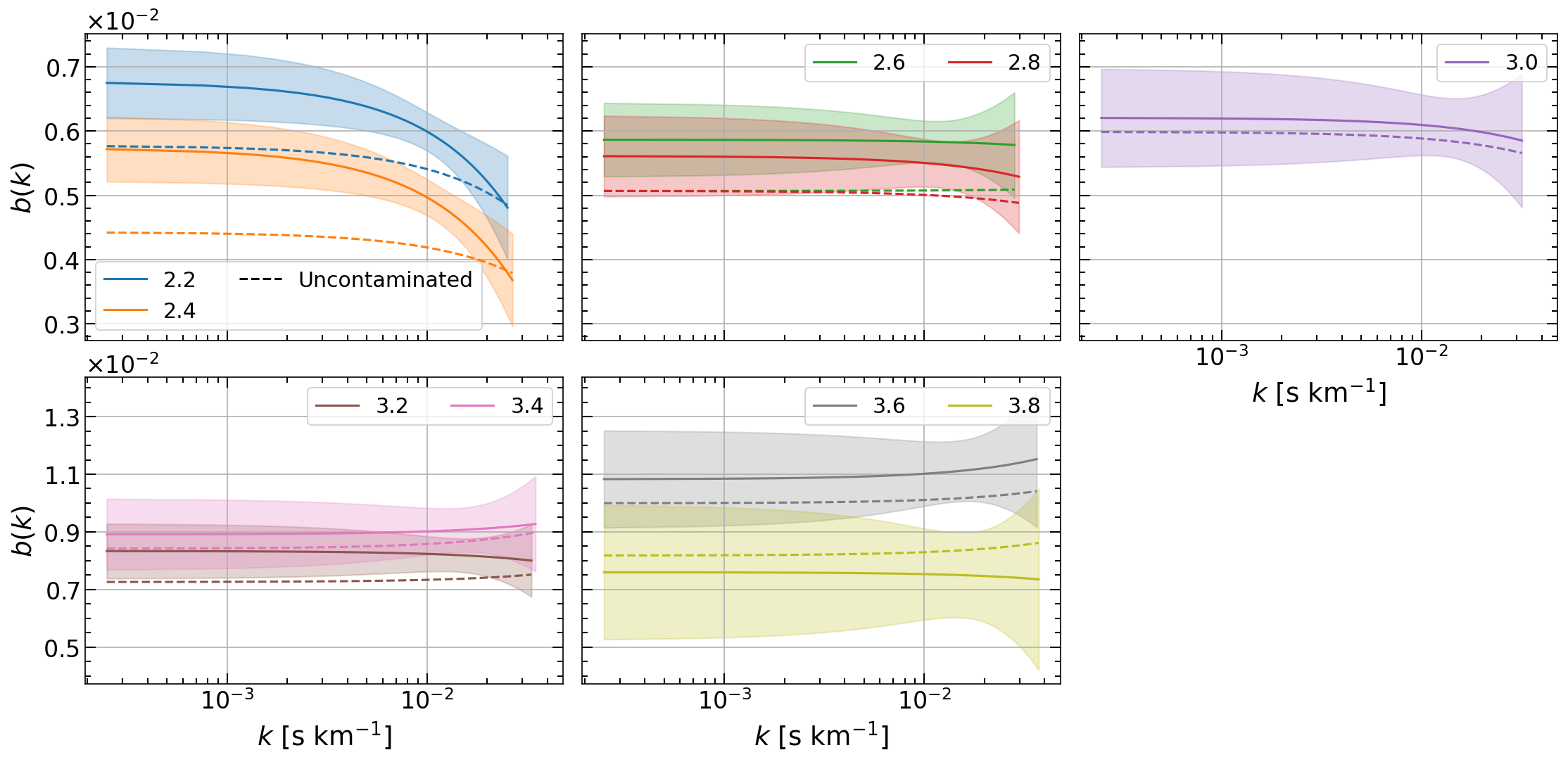}
    \caption{Fitted linear relation for $b(k)$ for redshifts bins with over 3$\sigma$ detection significance for the optimal estimator. The solid lines are derived from fully contaminated mocks, and the dashed lines are derived from uncontaminated mocks. The $b(k)$ values for the contaminated mocks are larger than those for the uncontaminated case for all redshifts except $z=3.8$. The difference arises from the continuum-fitting response to the underlying large-scale density field and the survey window function.
    }
    \label{fig:continuum_bias_corrections}
\end{figure}

Figure~\ref{fig:continuum_bias_corrections} shows the fitted polynomial and its $1\sigma$ contours for the first nine redshift bins where the bias is detected at greater than 3$\sigma$ significance. The solid lines and shaded regions denote $b(k)$ for the fully contaminated mock, whereas the dashed line is $b(k)$ derived from uncontaminated mocks discussed in section~\ref{subsec:qmle_val_nosyst}. It is important to note that the uncertainties reflect the statistical power of a 20 times larger dataset. The $b(k)$ values for the contaminated mocks are larger than those for the uncontaminated case for all redshifts except $z=3.8$. The difference is caused by the continuum-fitting response to the underlying large-scale density field within the unmasked forest region.
We conclude that the optimal estimator with this correction term (listed in table~\ref{tab:bias_corrections}) and a bootstrap covariance matrix is accurate.
\begin{table}
    \centering
    \begin{tabular}{|c|c || c | c || c | c|}
        \hline
        $z$ & $b(k) \times 10^3$ & $z$ & $b(k) \times 10^3$ & $z$ & $b(k) \times 10^3$ \\
        \hline
        2.2 & $6.768 - 0.699 x$ & 2.8 & $5.611 - 0.098 x$ & 3.4 & $8.913 + 0.093 x$ \\
        2.4 & $5.736 - 0.692 x$ & 3.0 & $6.206 - 0.101 x$ & 3.6 & $10.825 + 0.172 x$ \\
        2.6 & $5.865 - 0.026 x$ & 3.2 & $8.335 - 0.089 x$ & 3.8 & $7.599 - 0.059 x$ \\
        \hline
    \end{tabular}
    \caption{Bias corrections $b(k)$ for the optimal estimator method that are derived from the stack of 20 fully contaminated mocks, where $x \equiv k/0.009~$\skm.}
    \label{tab:bias_corrections}
\end{table}

\section{Validating the FFT estimator\label{sec:fft_val}}

In this section, we validate the FFT estimator. The KS test is performed over the same wavenumber range used in the EDR and DR1 measurements~\cite{ravouxFFTP1dEDR2023,ravouxFFTP1dDesiDr12024}, i.e., $0.0468 ~$\AA$^{-1} < k < 2.0$ \AA$^{-1}$. Since the small-number statistics in the higher redshift bins result in poor estimation of the covariance matrix and possible outliers, we limit the analysis to the redshift range $2.1 < z < 4.3$, as was done for the DR1 FFT measurement. 

The bias correction for FFT is different from the optimal estimator strategy. We fit a second-order polynomial to $P_\mathrm{ref} / P_m$ obtained from the stack of all realizations, where $m$ denotes the considered case, and correct each realization by directly multiplying it by this fit. Additionally, a systematic uncertainty is added to the covariance matrix used for $\chi^2$ calculation in the KS test. This uncertainty is defined as a percentage of the relative correction on \pk. For the previous EDR measurement~\cite{ravouxFFTP1dEDR2023}, the percentage was equal to 30\%. We use this percentage as the default for all the corrections tested in this article.\footnote{Except for BAL masking, where a smaller percentage is sufficient to pass the test as discussed later.} We incorporate this more realistic level of systematic uncertainty into all the cases below as part of the KS test.

\subsection{The power spectrum from transmission files}
The FFT estimator works in wavelength configuration instead of velocity configuration; that is, it estimates \poned\ in \AA\ units instead of \kms\ units. In this section, we test the consistency of the two.

The transformation between the two configurations is non-linear in general: $v = c \ln \lambda / \lambda_p$, such that the velocity separation between two points is $v_2 - v_1 = \Delta v = c\ln\lambda_2/\lambda_1$.  For small separations, this is approximated as $\Delta v = c \frac{\Delta\lambda}{\lambda_c}$, where $\Delta\lambda \equiv \lambda_2 - \lambda_1$ and $\lambda_c = (\lambda_1 + \lambda_2) / 2$. For a redshift bin $z$, the central wavelength becomes $\lambda_c = (1+z) \lambda_\mathrm{Ly\alpha}$. The conversion from \AA\ units to \kms\ units is then as follows:
\begin{equation}
    k_\lambda = \frac{c}{(1 + z) \lambda_\mathrm{Ly\alpha}} k_v \qquad\mathrm{and}\qquad P_\lambda = \frac{(1 + z) \lambda_\mathrm{Ly\alpha}}{c} P_v.
\end{equation}

\begin{figure}
    \centering
    \includegraphics[width=\linewidth]{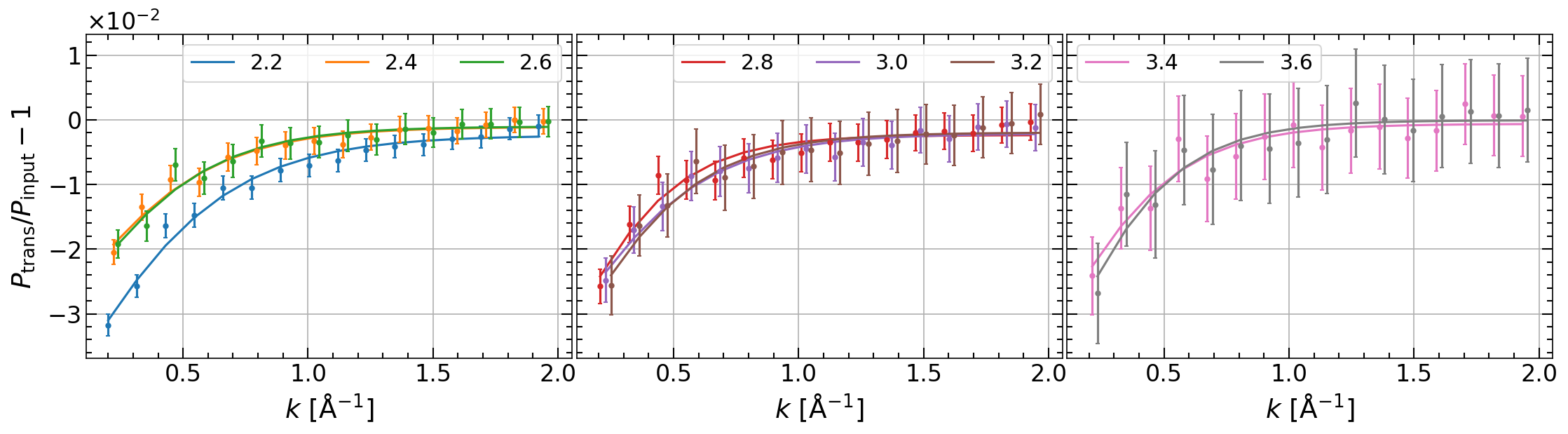}
    \caption{The relative difference between the power spectrum from mock transmissions ($P_\mathrm{trans}$) and the input power spectrum ($P_\mathrm{input}$) for the FFT estimator. The highest three redshift bins yield large error bars and are omitted for clarity. We find that the conversion between wavelength and velocity units introduces small biases at $k \leq 1~$\AA$^{-1}$ across all redshift bins as the small separations assumption gradually breaks down.}
    \label{fig:fft_raw_over_input}
\end{figure}

We compute the mock transmission power spectrum ($P_\mathrm{trans}$) and compare it to the input power spectrum ($P_\mathrm{input}$) converted from velocity to angstrom units to assess the validity range of this conversion in the absence of other complications. Figure~\ref{fig:fft_raw_over_input} shows the relative difference $P_\mathrm{trans} / P_\mathrm{input} - 1$. We omit the highest three redshift bins for clarity, since they yield large error bars. We find percent-level biases in wavenumbers below $k \leq 1~$\AA$^{-1}$ across all redshift bins. However, such a difference is not observed in the DR1 data between the optimal and FFT estimators, which may be obscured by other systematics in real data. 

This and other biases in the FFT estimator make it difficult to have a settled reference power spectrum for eq.~\eqref{eq:chi2}. In contrast to the optimal estimator analysis, the reference power spectrum is adapted to the considered case to mitigate effects other than the one being investigated.

\subsection{Validation in the absence of contaminants}

This section aims to (1) identify biases in the resolution treatment, (2) validate the covariance matrix, and (3) estimate the biases resulting from the continuum fitting procedure. As in section~\ref{subsec:qmle_val_nosyst}, we generate and use \numMocks\ \qq\ mocks without any contaminants for these goals.

\begin{figure}
    \centering
    \includegraphics[width=\columnwidth]{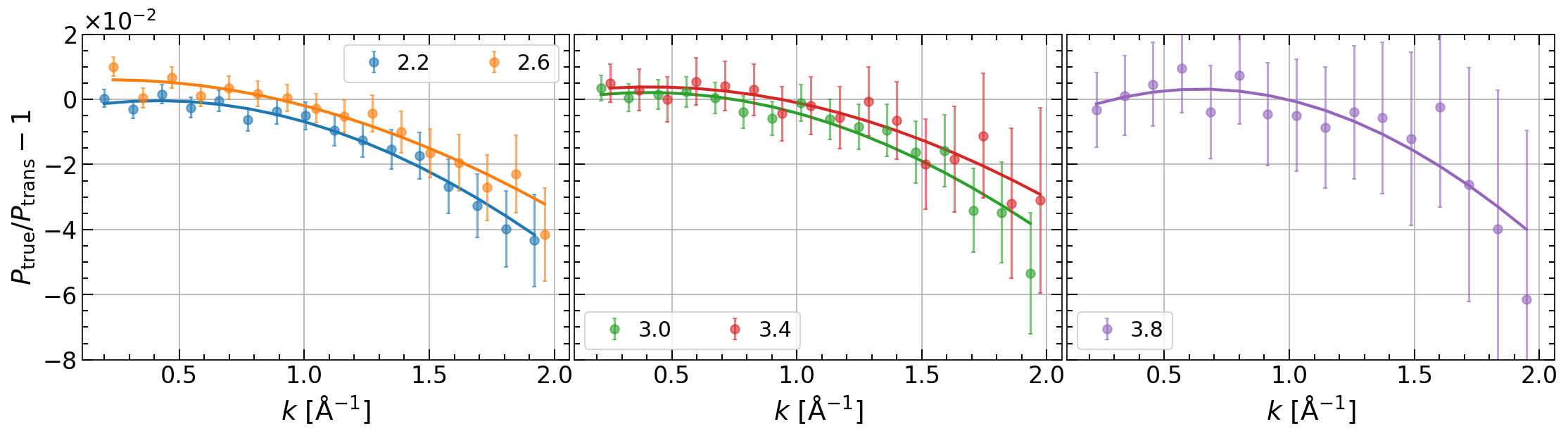}
    \caption{Comparison between the power spectrum measured on the transmission files and the one obtained using the true continuum imposed by \qq\ for the FFT estimator. Only a representative sample of redshift bins is shown for clarity.  This ratio quantifies the misestimation of the resolution damping correction, which affects the smallest scales. We use this ratio as an additional correction for the resolution in the DR1 measurement~\cite{ravouxFFTP1dDesiDr12024}. 
    }
    \label{fig:fft_raw_vs_truecont}
\end{figure}

We start by using the true continuum of quasars and mean flux as before. Using the above $P_\mathrm{trans}$ as the reference power spectrum, we investigate if any bias remains in the stack of all \numMocks\ mocks. Figure~\ref{fig:fft_raw_vs_truecont} shows the ratio between the two power spectra. We find a bias at small scales, which may be due to the Fourier treatment of the resolution matrix or an issue with noise subtraction. We remind the reader that \qq\ mocks do not involve complicated spectro-perfectionism reduction and instead provide an average resolution matrix for simulated cameras. We also note that the bias changes by $0.2\%$ on average across all redshifts (wavelengths), which indicates that this bias is due to the Fourier treatment of the resolution matrix.

\begin{figure}
    \centering
    \includegraphics[width=0.43\columnwidth]{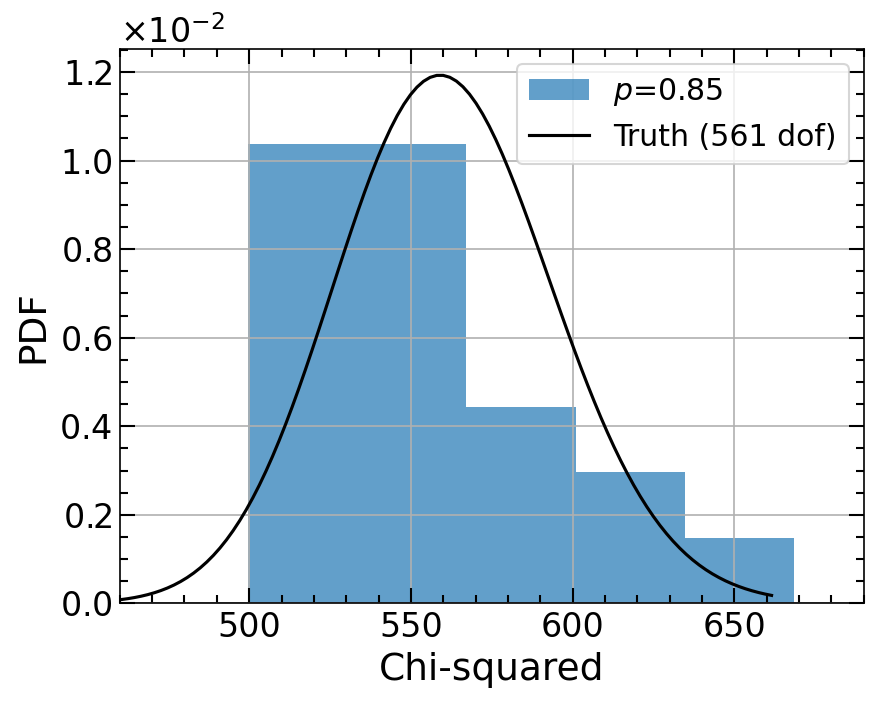}
    \includegraphics[width=0.43\columnwidth]{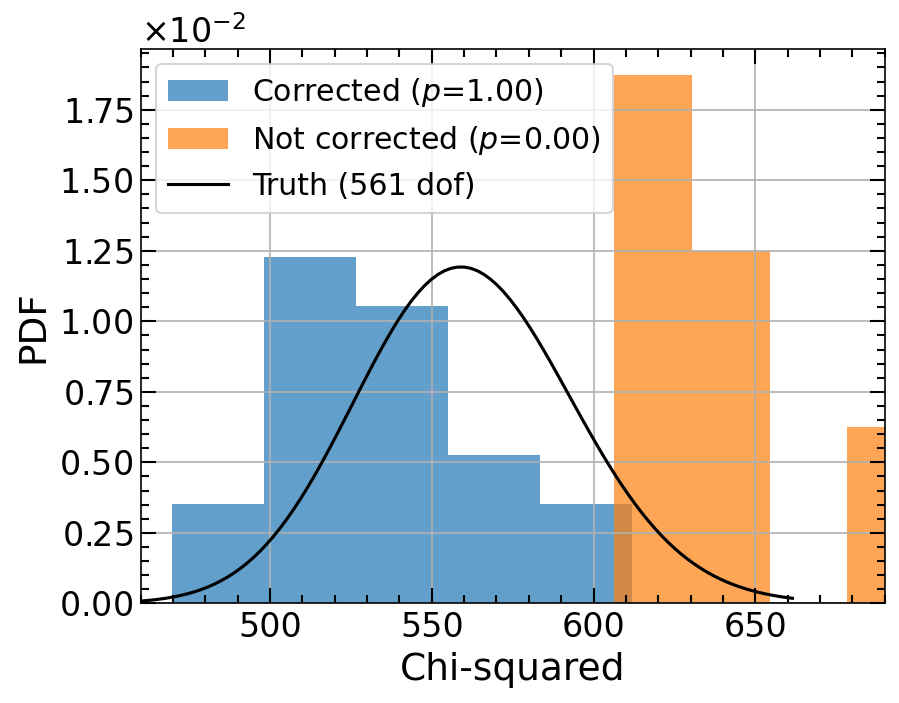}
    \caption{({\it Left}) PDF of chi-square values for the no systematics case for the FFT estimator. The one-sided KS test indicates that the covariance matrix estimator passes the validation threshold. ({\it Right}) PDF of $\chi^2$ values before ({\it orange}) and after ({\it blue}) continuum bias correction and resulting one-sided KS test p-value.}
    \label{fig:fft_test_covariance_truecont}
\end{figure}

To achieve our second goal, we set our reference power spectrum ($P_{\mathrm{ref}}$ in eq.~\eqref{eq:chi2}) as the stack of all \numMocks\ mocks. Since this reference power has the same resolution bias as the individual \poned\ estimates, the net contribution of resolution bias is removed from this test. To test the estimated covariance matrix of each realization, we compare the estimated \poned\ of this realization to the reference power spectrum and calculate $\chi^2$ using the covariance matrix derived for each realization (the errors in $P_{\mathrm{ref}}$ are ignored). Using the covariance from eq.~\eqref{FFT-cov} results in a one-sided p-value of one and a two-sided KS test p-value of zero. This indicates that the covariance is overestimated. We renormalized the covariance by a 10\% factor and show the resulting $\chi^2$ distribution in figure~\ref{fig:fft_test_covariance_truecont}. We obtain a one-sided p-value of 0.85 and a two-sided p-value of 0.27, which indicates that the $\chi^2$ distribution passes the KS test and is closer to the expected Gaussian distribution. We note that the relatively low value of the two-sided p-value is caused by the shape of the distribution rather than an amplitude offset in the covariance.

\begin{figure}
    \centering
    \includegraphics[width=\columnwidth]{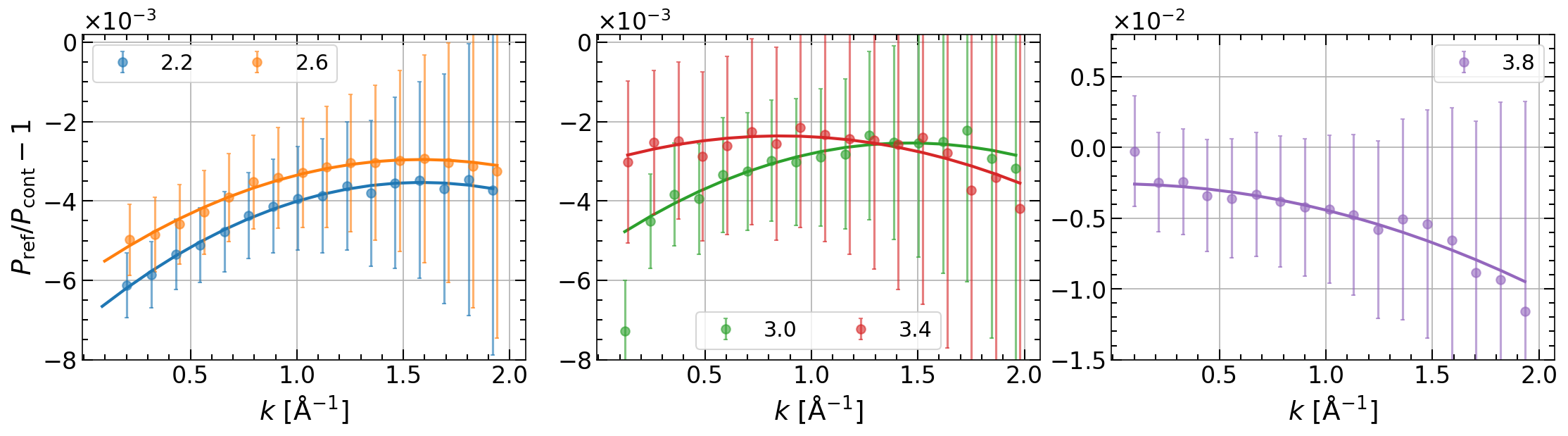}
    \caption{Impact of continuum fitting on \poned\ for the FFT estimator. Comparison between the average of 20 mocks with true continuum ($P_\mathrm{ref}$) and fitted continuum ($P_\mathrm{cont}$).  Only a representative sample of redshift bins is shown for clarity. We fit a second-order polynomial for $k > 0.1~$\AA\ that takes the error bar of each point into account.
    }
    \label{fig:fft_cont_corr}
\end{figure}

To estimate biases arising from continuum fitting, we use the same reference power spectrum, removing other possible causes that could explain the change in \poned\ while fitting the quasar continuum, such as spectrum pixelization, resolution modeling, and noise, since these affect both true continuum and fitted continuum estimates in the same manner. Figure~\ref{fig:fft_cont_corr} shows the impact of continuum fitting on \poned. This is a small effect (1\% at $z\leq3.2$ and up to 2\% at $z\geq 3.4$) and in agreement with conclusions from the optimal estimator and findings from EDR analysis in ref.~\cite{ravouxFFTP1dEDR2023}. 

We include a systematic error budget in the covariance matrix, as discussed before, to perform the KS test. 
The KS test with our correction and the additional error budget passes the p-value threshold, increasing it from zero for the uncorrected case to one, which is shown in the right panel of figure~\ref{fig:fft_test_covariance_truecont}. This is expected, considering that we are dealing with a relatively small effect.

\subsection{Deriving and validating masking corrections}

The FFT estimator requires data points to be regularly spaced. We mask pixels by setting the value of the \lya\ contrast $\delta_F$ to zero. Since all masked pixels are taken into account when computing \poned, masking introduces a bias that must be corrected. This bias is effectively a survey window function that convolves all Fourier modes. Here, we approximate this effect with a scale-dependent bias in Fourier space as was done in previous \lya\ \poned\ estimates with FFT. Our baseline approach follows refs.~\cite{chabanierOnedimensionalPowerSpectrum2019, ravouxFFTP1dEDR2023}, i.e.\ we use the FFT estimator to compute the ratio between the masked and unmasked power spectra of \qq\ mocks. For each redshift bin, this ratio is fit with an appropriate polynomial to derive a correction that we will multiplicatively apply to the data in a later step. In this section, the masking corrections for HCD, atmospheric lines, and BAL are computed and tested. Additionally, we test the multiplicativity of these masking corrections when they are combined with the continuum fitting correction defined in previous sections.

\subsubsection{HCD masking}
The presence of non-IGM components due to HCDs, in addition to the \lya\ forest absorption due to neutral hydrogen in the IGM, significantly impacts the measured \poned. 
These HCDs produce deep and wide absorption in the \lya\ forest with extended damping wings. These features occur when the sightline to a quasar passes through the circumgalactic environment of an intervening galaxy. The contamination by these systems is a major source of systematic uncertainty in the estimation and cosmological interpretation of \poned, especially for small values of $k$ (e.g., see ref.~\cite{rogersEffectsOfHcd2018}).

To assess the impact of these effects, as well as the overall impact on \poned\, we rely again on \numMocks\ \qq\ mocks. The effect of pure masking is tested by masking the DLA regions using their exact locations and column densities in \emph{uncontaminated} mocks. This isolates the effect solely to masking by removing the complications that come from a mismatch in the sizes of DLA regions estimated using column densities. We perform continuum fitting correction with and without masking HCDs and compare their ratio. The resulting bias is below the percent level and nearly featureless (constant in $k$). Nevertheless, when uncorrected, the estimates yield a p-value of zero, failing the KS test. We pass the KS test with a p-value of one by applying the HCD correction. 

\subsubsection{Discrete line masking}
We mask the atmospheric emission and Galactic absorption lines identified in ref.~\cite{ravouxFFTP1dEDR2023}. These features occur at fixed observed wavelengths across all spectra, so that they introduce a larger bias through these correlations.

\begin{figure}
    \centering
    \includegraphics[width=\columnwidth]{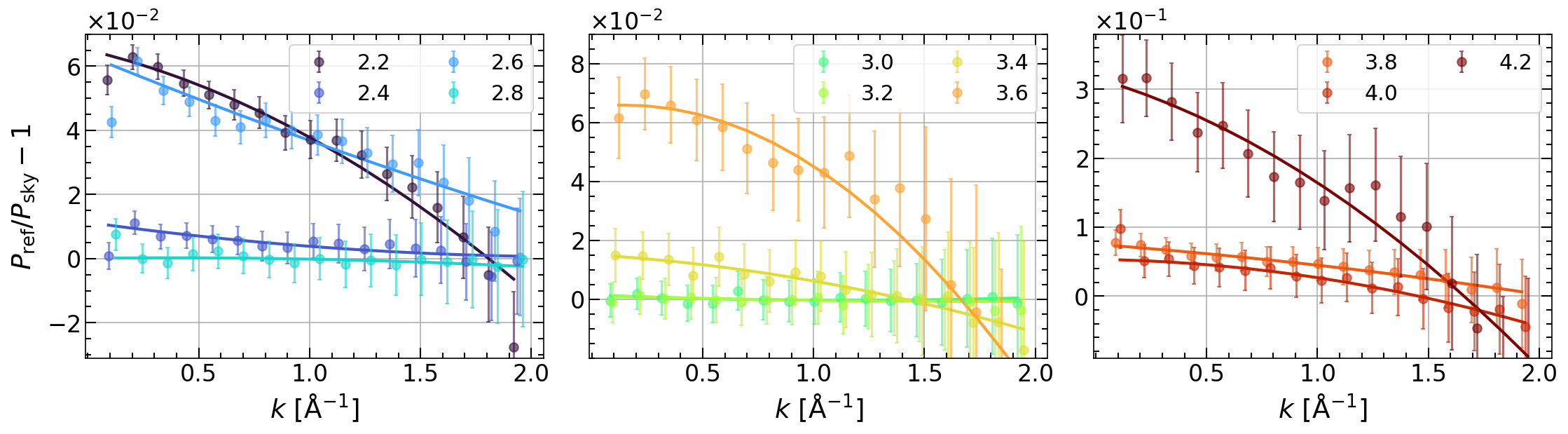}
    \caption{Impact of atmospheric emission line masking on the one-dimensional power spectrum for the FFT estimator. The impact is quantified by the ratio of the reference power spectrum $(P_\mathrm{ref})$
    to the atmospheric emission line masked power spectrum $(P_\mathrm{sky})$. The fitted solid lines are used as corrections for the bias induced by atmospheric emission line masking.
    }
    \label{fig:fft_atmline_corr}
\end{figure}

In this section, we improve the precision of our bias correction using the new, larger set of mocks tailored for DESI DR1. As before, we calculate the ratios $P_{\rm sky}/P_{\rm ref}$ and fit a second-order polynomial to the stack of all realizations. These are shown in figure~\ref{fig:fft_atmline_corr}. The largest corrections occur at redshift bins $z=2.2$ and $z=2.6$ due to sparse atmospheric line emissions and Galactic absorption due to the Ca~\textsc{ii} H\&K doublet lines at 3968~\AA\ and 3933~\AA. For $z\geq 2.8$, the corrections are only caused by atmospheric emission line masking.

The left panel of figure~\ref{fig:fft_multiplicativity_corr} shows the $\chi^2$ distribution for the KS test, which gives a p-value of 0.90. This test indicates that the correction derived is well-suited, even with a significant impact on the power spectrum. It also indicates that the systematic uncertainty associated with this correction (which is set to 30~\% of the correction itself~\cite{ravouxFFTP1dDesiDr12024}, included in the covariance matrix) is properly estimated.

\subsubsection{BAL masking}

\begin{figure}
    \centering
    \includegraphics[width=\columnwidth]{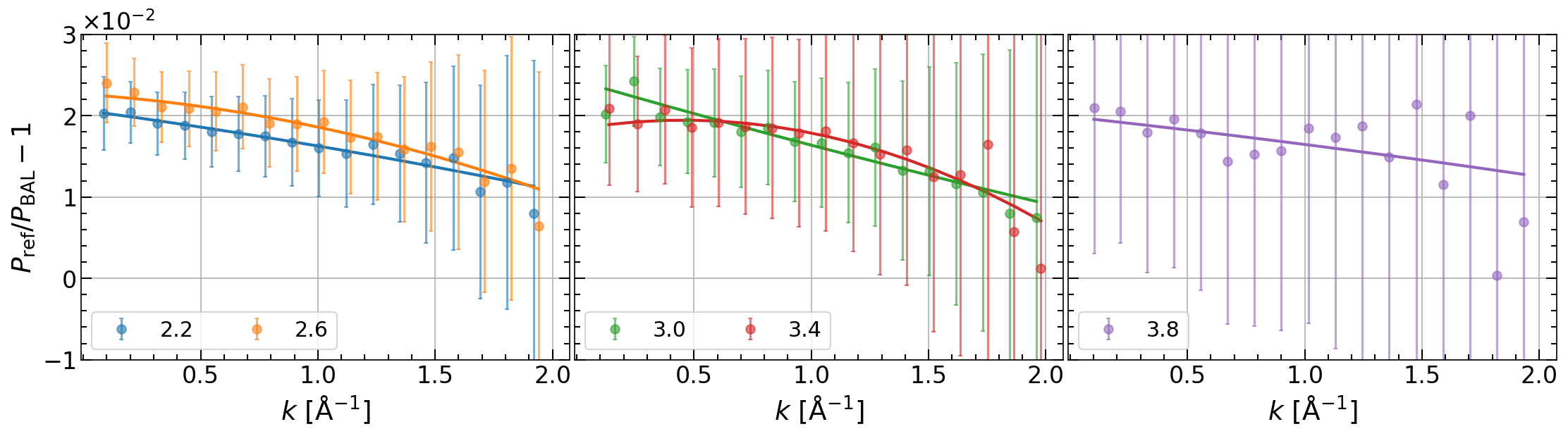}
    \caption{Impact of BAL $AI$ masking on the one-dimensional power spectrum for the FFT estimator. The impact is quantified by the ratio of the reference power spectrum $(P_\mathrm{ref})$ to the BAL $AI$ masked power spectrum $(P_\mathrm{BAL})$.  Only a representative sample of redshift bins is shown for clarity.  The fitted solid lines are used as corrections for the bias induced by BAL $AI$ masking.
    }
    \label{fig:fft_balm_corr}
\end{figure}

The companion paper \cite{ravouxFFTP1dDesiDr12024} performs a variation of the FFT analysis that shows that BALs detected with Absorption Index ($AI$, see ref.~\cite{filbertBALEDRcatalog2023} for details) cause a significant deviation at large scales and should be accounted for. Here, we derive a correction associated with the masking of BAL systems with $AI > 0$ by comparing \poned\ when features are masked to the unmasked case. This is shown in figure~\ref{fig:fft_balm_corr}.

For the KS test in this case, we decrease the associated relative systematic uncertainty caused by the BAL $AI$ masking correction to 6\%, as the default value (30\%) is a significant overestimate according to the $\chi^2$ distribution. With this new uncertainty, we consider the obtained one-sided p-value of 0.45 (shown in the left panel of figure~\ref{fig:fft_multiplicativity_corr}) sufficient to validate the correction and its systematic uncertainty.

\subsubsection{Combining corrections}

\begin{figure}
    \centering
    \includegraphics[width=0.8\columnwidth]{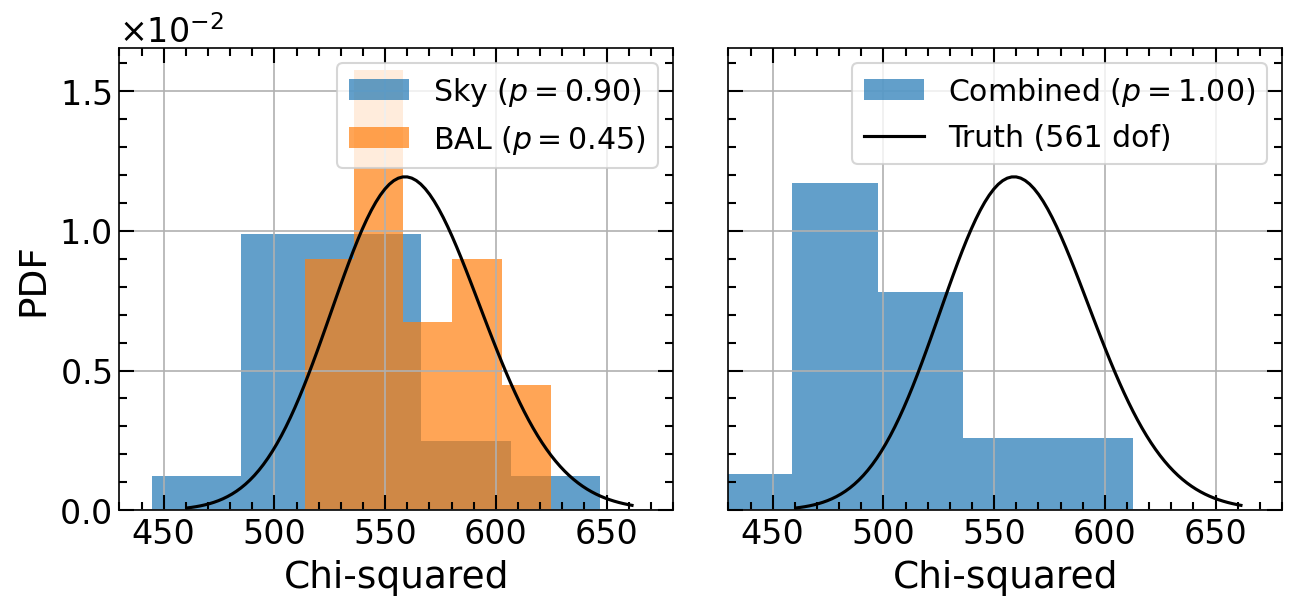}
    \caption{({\it Left}) PDFs of $\chi^2$ values for the FFT estimator after atmospheric (sky) line masking correction ({\it blue}) and BAL $AI$ masking correction ({\it orange}). The KS test indicates that the covariance matrix estimator with added systematic error budget passes the validation threshold. ({\it Right}) PDF of $\chi^2$ values after the combined effect of DLA, BAL $AI$, and sky masking is corrected by multiplication of individual effects. The KS test shows that these corrections can be multiplied.}
    \label{fig:fft_multiplicativity_corr}
\end{figure}

The FFT estimator requires several corrections associated with continuum fitting, DLA, BAL $AI$, and atmospheric line masking. The main measurement assumes the multiplicativity of these corrections, i.e., if all effects are applied simultaneously, the full correction is the product of individual corrections. However, this assumption does not account for potential correlations between the effects. Masked regions of different effects can significantly overlap, resulting in an overestimation when the product of corrections is applied. Additionally,  some masking corrections can differ with and without continuum fitting. 

In order to test the multiplicativity assumption, we create a set of mocks with continuum fitting for which the three masking effects are applied. We correct the \pk~of each mock by the product of individual corrections. Furthermore, we include a systematic uncertainty that is the quadratic sum of the previously described individual uncertainties. The corrected measurement is compared to the one using the true continuum of quasars, and the results of the KS test are shown in the right panel of figure~\ref{fig:fft_multiplicativity_corr}. Without changing the percentage associated with 
systematics, the one-sided p-value of one indicates that the assumption of correction multiplicativity is verified as correct. The lower $\chi^2$ values compared to expectation indicate that the whole combination of systematic uncertainties is likely overestimated.

\section{Accuracy of the spectrograph resolution\label{sec:ccd_resolution_val}}
Uncertainty in the resolution correction dominates the systematic error budget for \poned\ at large wavenumbers. When the bias in the spectrograph resolution $(b_\mathrm{res})$ is small, the scale-dependent error on \poned\ can be written as $b_\mathrm{res} \times 2k^2 R_z^2 \times P_\mathrm{true}(k, z)$, where $R_z \equiv c \Delta\lambda_\mathrm{DESI} / (1 + z) \lambda_\mathrm{Ly\alpha}$ \cite{karacayliOptimal1dDesiEdr2023}.
Our goal is to quantify these systematic errors on \poned\ using CCD image simulations processed with the DESI spectroscopic pipeline (described in section~\ref{subsec:ccd_image_sims}). In a previous work, we validated the spectro-perfectionism algorithm and its resolution matrix in \poned\ estimation when the PSF is perfectly known \cite{karacayliOptimal1dDesiEdr2023}. We also showed that there is a wavelength-dependent bias of $b_\mathrm{res}(z) \sim 1\%$ in the spectrograph resolution when the PSF is empirically measured from arc calibration lamps. Our focus here is on a better quantification of $b_\mathrm{res}(z)$.

\begin{figure}
    \centering
    \includegraphics[width=\linewidth]{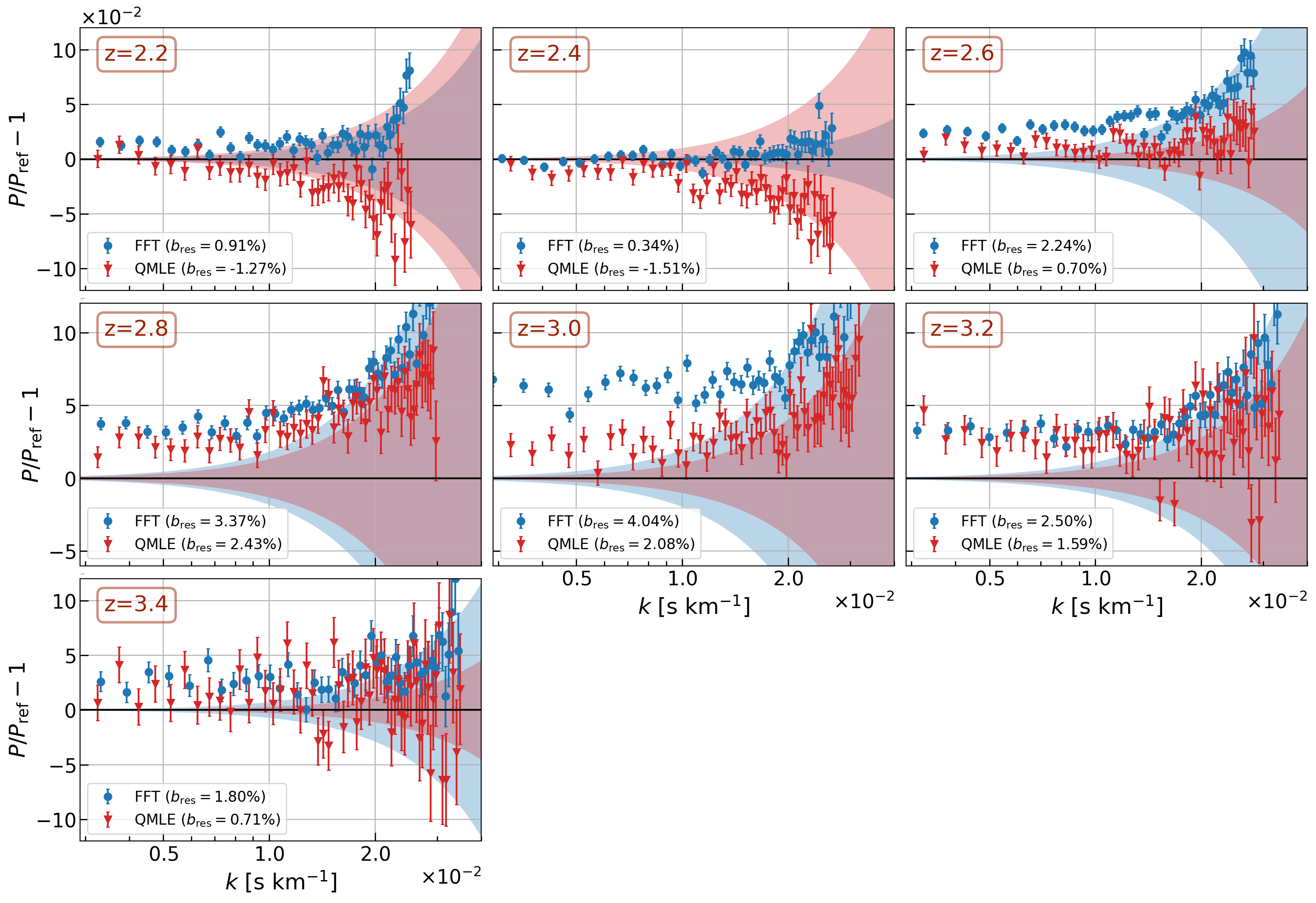}
    \caption{Systematic errors due to the spectral resolution derived from CCD image simulations for both estimators. As we have shown previously in ref.~\cite{karacayliOptimal1dDesiEdr2023} for the optimal estimator, the bias depends on the wavelength, changes direction, and corresponds to $1-2\%$ precision. The bias of the FFT estimator seems consistent in direction across wavelengths and agrees with the optimal estimator results except for $z=3.0$. We do not correct for this bias but instead include it in the systematic error budget as a correlated fluctuation.}
    \label{fig:qmle_ccd_systerr}
\end{figure}

We simulate 675,000 spectra, which is fifteen times the number used in our EDR analysis, and extract them using empirical PSFs in the same manner as DESI's nightly operations, using the same version of the pipeline used for DR1 processing. The relative errors of both estimators with respect to the reference power spectrum (analytically calculated for the optimal estimator and measured from the transmission files for the FFT estimator) are shown in figure~\ref{fig:qmle_ccd_systerr}, where the blue points stand for the FFT results and the red triangles for the optimal estimator results. We remind the reader that we generated these simulations with a maximum quasar redshift of $z_\mathrm{qso}^\mathrm{max}=3.6$, corresponding to a maximum forest redshift bin of $z^\mathrm{max}_F=3.4$.
The two estimators generally display the same trend in error, except for a few key differences. First, they show different error trends at high $k$ for $z=2.2$ and $2.4$. Such differences are normal, as each estimator has a unique treatment for the resolution correction. Second, the FFT estimator has a larger bias at low $k$ than the optimal estimator. This large-scale bias is unexpected and is unlikely to originate from a purely resolution effect.  We do not include this large-scale offset or remove its contribution at high $k$ by keeping the $k^2$ functional form without additive terms. The $b_\mathrm{res}(z)$ are derived in each redshift bin independently and noted in the figure.

Our recommendation is to include these biases in the systematic error budget as correlated fluctuations, rather than correcting for them in the measurement. First, the simulation fidelity needs to be improved, and the large-scale offset must be understood to ensure the precision of such corrections. Second, these biases partially arise from errors in estimating pipeline noise. Later versions of the spectroscopic pipeline include minor changes in this area. We reprocess the simulated spectra with the latest version of the pipeline available and confirm that these changes alter, but do not eliminate, the high $k$ bias. This is an ongoing effort.

The offset at large scales remains in both versions of the spectroscopic pipeline. Various diagnostic quantities do not indicate any problems with the pipeline, such as stacked flux for measuring flux calibration errors and estimating power in spectral regions with no power (null test). We also do not observe a 5\% difference in real data between the two estimators at $z=3.0$, which makes us assume that this is an error in the simulation stage rather than the extraction stage. This will also be investigated in the future.

\section{Discussion\label{sec:discuss}}
We have validated the \poned\ measurement pipelines for DESI DR1 against major systematics such as masking contaminated pixels (survey window function), biases arising from continuum fitting errors, and the spectroscopic pipeline for both the optimal and FFT estimators. We generated two classes of synthetic spectra.

The first class is based on 1D \lya\ transmission fields and includes a diverse set of quasar spectra and simulates the DESI instrument in 1D. We generated 20 realizations of DR1 and used these spectra to (1) validate that the optimal estimator is robust against masked pixels and (2) derive masking bias corrections for the FFT estimator and validate their application. Additionally, we calculated the continuum fitting error bias for both estimators.

The second class of synthetic spectra projects quasar spectra onto realistic, 2D CCD images that simulate raw DESI data. These are then processed with the DESI spectroscopic reduction pipeline. Using these CCD image simulations, we quantified the accuracy of the spectrograph resolution to be around 1\%. The result of all of these tests is that the DESI \poned\ measurement pipeline is robust and understood at the percent level.

Given the increased number and signal-to-noise ratio of spectra from ongoing DESI observations, it is important to continue improving synthetic data generation to take full advantage of the data's increased statistical power. First, the mock power spectrum is only tuned to match the observations with 10--20\% agreement on average. Improving the fidelity of the mocks to the data will improve the accuracy of the various corrections presented in this article, and enable end-to-end \poned\ validation analyses that include cosmological inference in future pipelines.

Relatedly, HCDs are randomly placed into our mocks, and their column density distribution is based on measurements from smaller samples than are currently available. The masked regions at HCD locations are correlated with the \lya\ forest, which introduces an additional bias for both estimators in real data. In a future study, these systems can be placed with an abundance matching formalism to include forest-HCD correlations and study the cosmological implications of HCDs beyond masked pixels. This 1D formalism would be significantly cheaper than realistic hydrodynamical simulations. Furthermore, in real data analysis, only the densest systems, DLAs with column density $N_{\rm{HI}}>10^{20.3}~{\rm cm^{-2}}$, can be identified with 85\% completeness and purity depending on the confidence level and SNR cuts \cite{wangDeepLearningDESIDLA2022, mingfengDLAGP2021, brodzellerConstructionDlaDesiDr2}. This means all weak HCDs ($N_{\rm HI}<10^{20.3}~{\rm cm^{-2}}$), as well as missed DLAs, are left unmasked in the data. They impact \poned\ through their intrinsic average line profile (one-halo effect) and their clustering (two-halo effect), since they are biased tracers of cosmic structure. For similar reasons, falsely detected DLAs may also bias the estimated \poned. As the DLA finder is expected to find false positives preferentially in regions of large absorption, the associated mask is correlated with the \lya\ forest. We leave the study of these effects to future work.

None of our simulations includes masking due to cosmic rays. However, the statistics of such pixels are minor and are not expected to alter the measurement at levels comparable to those of atmospheric lines or other astrophysical contaminants. Finally, we note that FFT masking can be improved by applying analytic window correction methods, as in ref.~\cite{Lokken2025}, thereby drastically reducing the systematic error introduced by masking. This improvement will be performed in future studies.

\section*{Data Availability}
All data points shown in the figures will be available after publication.

\paragraph{Software.} The optimal estimator\footnote{\url{https://github.com/p-slash/lyspeq}} is written in \texttt{c++}.
It depends on \texttt{CBLAS} and \texttt{LAPACKE} routines for matrix/vector operations, \texttt{GSL}\footnote{\url{https://www.gnu.org/software/gsl}} for certain scientific calculations \citep{GSL}, \texttt{FFTW3}\footnote{\url{https://fftw.org}} for FFTs \citep{FFTW05}; and uses the Message Passing Interface (MPI) standard\footnote{\url{https://www.mpich.org}} to parallelize tasks. The FFT estimator and associated plots are generated with \texttt{picca}$^{\ref{footnote:picca}}$ (9.12.3) and \texttt{p1desi}\footnote{\url{https://github.com/corentinravoux/p1desi}} (2.0.1) software. 
The quasar spectra are organized with the \texttt{HEALPix} \citep{healpix} scheme on the sky.
We use the following commonly-used software in \texttt{python} analysis: \texttt{astropy}\footnote{\url{https://www.astropy.org}}
a community-developed core \texttt{python} package for Astronomy \citep{astropy:2013, astropy:2018, astropy:2022},
\texttt{numpy}\footnote{\url{https://numpy.org}}
an open source project aiming to enable numerical computing with \texttt{python} \citep{numpy},
\texttt{scipy}\footnote{\url{https://scipy.org}} an open-source project with algorithms for scientific computing.
\texttt{healpy} to interface with \texttt{HEALPix} in \texttt{python} \citep{healpy},
Finally, we make plots using
\texttt{matplotlib}\footnote{\url{https://matplotlib.org}}
a comprehensive library for creating static, animated, and interactive visualizations in \texttt{python}
\citep{matplotlib}.

\acknowledgments

NGK and PM acknowledge support from the United States Department of Energy, Office of High Energy Physics under Award Number DE-SC0011726. EA acknowledges support from the Agence Nationale de la Recherche (ANR) through grant ANR-22-CE92-0037.

This material is based upon work supported by the U.S. Department of Energy (DOE), Office of Science, Office of High-Energy Physics, under Contract No. DE–AC02–05CH11231, and by the National Energy Research Scientific Computing Center, a DOE Office of Science User Facility under the same contract. Additional support for DESI was provided by the U.S. National Science Foundation (NSF), Division of Astronomical Sciences under Contract No. AST-0950945 to the NSF’s National Optical-Infrared Astronomy Research Laboratory; the Science and Technology Facilities Council of the United Kingdom; the Gordon and Betty Moore Foundation; the Heising-Simons Foundation; the French Alternative Energies and Atomic Energy Commission (CEA); the National Council of Humanities, Science and Technology of Mexico (CONAHCYT); the Ministry of Science, Innovation and Universities of Spain (MICIU/AEI/10.13039/501100011033), and by the DESI Member Institutions: \url{https://www.desi.lbl.gov/collaborating-institutions}. Any opinions, findings, and conclusions or recommendations expressed in this material are those of the author(s) and do not necessarily reflect the views of the U. S. National Science Foundation, the U. S. Department of Energy, or any of the listed funding agencies.

The authors are honored to be permitted to conduct scientific research on I'oligam Du'ag (Kitt Peak), a mountain with particular significance to the Tohono O’odham Nation.

\appendix

\section{Author Affiliations}
\label{sec:affiliations}






\noindent \hangindent=.5cm $^{6}${Lawrence Berkeley National Laboratory, 1 Cyclotron Road, Berkeley, CA 94720, USA}

\noindent \hangindent=.5cm $^{7}${Department of Physics, Boston University, 590 Commonwealth Avenue, Boston, MA 02215 USA}

\noindent \hangindent=.5cm $^{8}${Department of Physics \& Astronomy, University of Rochester, 206 Bausch and Lomb Hall, P.O. Box 270171, Rochester, NY 14627-0171, USA}

\noindent \hangindent=.5cm $^{9}${Dipartimento di Fisica ``Aldo Pontremoli'', Universit\`a degli Studi di Milano, Via Celoria 16, I-20133 Milano, Italy}

\noindent \hangindent=.5cm $^{10}${INAF-Osservatorio Astronomico di Brera, Via Brera 28, 20122 Milano, Italy}

\noindent \hangindent=.5cm $^{11}${Department of Physics \& Astronomy, University College London, Gower Street, London, WC1E 6BT, UK}

\noindent \hangindent=.5cm $^{12}${NASA Einstein Fellow}

\noindent \hangindent=.5cm $^{13}${Instituto de F\'{\i}sica, Universidad Nacional Aut\'{o}noma de M\'{e}xico,  Circuito de la Investigaci\'{o}n Cient\'{\i}fica, Ciudad Universitaria, Cd. de M\'{e}xico  C.~P.~04510,  M\'{e}xico}

\noindent \hangindent=.5cm $^{14}${Department of Astronomy \& Astrophysics, University of Toronto, Toronto, ON M5S 3H4, Canada}

\noindent \hangindent=.5cm $^{15}${Department of Physics \& Astronomy and Pittsburgh Particle Physics, Astrophysics, and Cosmology Center (PITT PACC), University of Pittsburgh, 3941 O'Hara Street, Pittsburgh, PA 15260, USA}

\noindent \hangindent=.5cm $^{16}${University of California, Berkeley, 110 Sproul Hall \#5800 Berkeley, CA 94720, USA}

\noindent \hangindent=.5cm $^{17}${Institut de F\'{i}sica d’Altes Energies (IFAE), The Barcelona Institute of Science and Technology, Edifici Cn, Campus UAB, 08193, Bellaterra (Barcelona), Spain}

\noindent \hangindent=.5cm $^{18}${Departamento de F\'isica, Universidad de los Andes, Cra. 1 No. 18A-10, Edificio Ip, CP 111711, Bogot\'a, Colombia}

\noindent \hangindent=.5cm $^{19}${Observatorio Astron\'omico, Universidad de los Andes, Cra. 1 No. 18A-10, Edificio H, CP 111711 Bogot\'a, Colombia}

\noindent \hangindent=.5cm $^{20}${Institut d'Estudis Espacials de Catalunya (IEEC), c/ Esteve Terradas 1, Edifici RDIT, Campus PMT-UPC, 08860 Castelldefels, Spain}

\noindent \hangindent=.5cm $^{21}${Institute of Cosmology and Gravitation, University of Portsmouth, Dennis Sciama Building, Portsmouth, PO1 3FX, UK}

\noindent \hangindent=.5cm $^{22}${Institute of Space Sciences, ICE-CSIC, Campus UAB, Carrer de Can Magrans s/n, 08913 Bellaterra, Barcelona, Spain}

\noindent \hangindent=.5cm $^{23}${University of Virginia, Department of Astronomy, Charlottesville, VA 22904, USA}

\noindent \hangindent=.5cm $^{24}${Fermi National Accelerator Laboratory, PO Box 500, Batavia, IL 60510, USA}

\noindent \hangindent=.5cm $^{25}${Institut d'Astrophysique de Paris. 98 bis boulevard Arago. 75014 Paris, France}

\noindent \hangindent=.5cm $^{26}${Department of Physics, The University of Texas at Dallas, 800 W. Campbell Rd., Richardson, TX 75080, USA}

\noindent \hangindent=.5cm $^{27}${NSF NOIRLab, 950 N. Cherry Ave., Tucson, AZ 85719, USA}

\noindent \hangindent=.5cm $^{28}${Department of Physics and Astronomy, University of California, Irvine, 92697, USA}

\noindent \hangindent=.5cm $^{29}${Sorbonne Universit\'{e}, CNRS/IN2P3, Laboratoire de Physique Nucl\'{e}aire et de Hautes Energies (LPNHE), FR-75005 Paris, France}

\noindent \hangindent=.5cm $^{30}${Departament de F\'{i}sica, Serra H\'{u}nter, Universitat Aut\`{o}noma de Barcelona, 08193 Bellaterra (Barcelona), Spain}

\noindent \hangindent=.5cm $^{31}${Instituci\'{o} Catalana de Recerca i Estudis Avan\c{c}ats, Passeig de Llu\'{\i}s Companys, 23, 08010 Barcelona, Spain}

\noindent \hangindent=.5cm $^{32}${Departamento de F\'{\i}sica, DCI-Campus Le\'{o}n, Universidad de Guanajuato, Loma del Bosque 103, Le\'{o}n, Guanajuato C.~P.~37150, M\'{e}xico}

\noindent \hangindent=.5cm $^{33}${Instituto Avanzado de Cosmolog\'{\i}a A.~C., San Marcos 11 - Atenas 202. Magdalena Contreras. Ciudad de M\'{e}xico C.~P.~10720, M\'{e}xico}

\noindent \hangindent=.5cm $^{34}${Department of Physics and Astronomy, University of Waterloo, 200 University Ave W, Waterloo, ON N2L 3G1, Canada}

\noindent \hangindent=.5cm $^{35}${Perimeter Institute for Theoretical Physics, 31 Caroline St. North, Waterloo, ON N2L 2Y5, Canada}

\noindent \hangindent=.5cm $^{36}${Waterloo Centre for Astrophysics, University of Waterloo, 200 University Ave W, Waterloo, ON N2L 3G1, Canada}

\noindent \hangindent=.5cm $^{37}${Space Sciences Laboratory, University of California, Berkeley, 7 Gauss Way, Berkeley, CA  94720, USA}

\noindent \hangindent=.5cm $^{38}${Instituto de Astrof\'{i}sica de Andaluc\'{i}a (CSIC), Glorieta de la Astronom\'{i}a, s/n, E-18008 Granada, Spain}

\noindent \hangindent=.5cm $^{39}${Departament de F\'isica, EEBE, Universitat Polit\`ecnica de Catalunya, c/Eduard Maristany 10, 08930 Barcelona, Spain}

\noindent \hangindent=.5cm $^{40}${Department of Physics and Astronomy, Sejong University, 209 Neungdong-ro, Gwangjin-gu, Seoul 05006, Republic of Korea}

\noindent \hangindent=.5cm $^{41}${CIEMAT, Avenida Complutense 40, E-28040 Madrid, Spain}

\noindent \hangindent=.5cm $^{42}${Department of Physics, University of Michigan, 450 Church Street, Ann Arbor, MI 48109, USA}

\noindent \hangindent=.5cm $^{43}${University of Michigan, 500 S. State Street, Ann Arbor, MI 48109, USA}

\noindent \hangindent=.5cm $^{44}${Department of Physics \& Astronomy, Ohio University, 139 University Terrace, Athens, OH 45701, USA}

\noindent \hangindent=.5cm $^{45}${Excellence Cluster ORIGINS, Boltzmannstrasse 2, D-85748 Garching, Germany}

\noindent \hangindent=.5cm $^{46}${University Observatory, Faculty of Physics, Ludwig-Maximilians-Universit\"{a}t, Scheinerstr. 1, 81677 M\"{u}nchen, Germany}

\noindent \hangindent=.5cm $^{47}${National Astronomical Observatories, Chinese Academy of Sciences, A20 Datun Road, Chaoyang District, Beijing, 100101, P.~R.~China}

\bibliographystyle{JHEP}
\bibliography{references}

\end{document}